\documentclass[preprint, 3p,12pt,sort&compress]{elsarticle}
\journal{Advanced Engineering Informatics}
\usepackage[utf8]{inputenc}
\usepackage{booktabs}
\usepackage{amsmath}

\usepackage{mathrsfs}  
\usepackage{empheq}
\usepackage{amssymb}
\usepackage{graphicx}
\usepackage{diagbox} 
\usepackage{derivative}
\usepackage{multicol}
\usepackage{multirow}
\usepackage{pdfpages}
\DeclareSymbolFont{rsfs}{U}{rsfs}{m}{n}
\DeclareSymbolFontAlphabet{\mathscrsfs}{rsfs}
\usepackage{comment}
\usepackage{tabularx}
\usepackage{tabularray}
\graphicspath{ {./images/} }
\usepackage{float,booktabs,array}
\usepackage[pagewise]{lineno}
\usepackage[ruled,vlined,linesnumbered,algo2e,resetcount]{algorithm2e}
\usepackage{algorithm}
\usepackage{algcompatible}
\usepackage{lineno}
\usepackage{xcolor}
\usepackage{tikz,pgf} 
\usepackage{color, soul}
\usepackage{siunitx}
\usepackage{subcaption}
\usepackage[labelsep=period]{caption}
\usepackage[colorlinks]{hyperref}
\usepackage{enumitem}
\usepackage{setspace}
\usepackage{arydshln}

\newcolumntype{Y}{>{\centering\arraybackslash}X}
\usepackage[flushleft]{threeparttable}
\usepackage{xltabular}
\usepackage[bottom]{footmisc}
\usetikzlibrary{arrows.meta}
\definecolor{green}{rgb}{0.0,0.50,0.0}
\tikzset{>={Straight Barb[angle'=80, scale=1.1]}}
\hypersetup{
     colorlinks=true,
     linkcolor=blue,
     filecolor=blue,
     citecolor = blue,      
     urlcolor=blue,
     pdfauthor=author
     }

\usepackage{framed}
\usepackage{nomencl}
\usepackage{longtable}
\makenomenclature
\usepackage{etoolbox}

\usepackage{multicol}
\usepackage{hyperref}
\renewcommand{\nomgroup}[1]{}
\usepackage{framed}
\renewcommand*\nompreamble{\begin{multicols}{2}}
\renewcommand*\nompostamble{\end{multicols}}
\makenomenclature

\begin{document}

\begin{frontmatter}

\title{Transformer-based Multisensor Data Fusion of Ultrasonic Guided Wave and FBG-based Strain Measurements for Multitask Aerospace Structural Health Monitoring}

\author[1]{Xin Yang\corref{correspondingauthor}}
\ead{xin.yang@kuleuven.be}
\author[2]{Morteza Moradi}
\author[3]{Tongtong Yan}
\author[4]{Jinbo Du}
\author[5]{Yunlai Liao}
\author[2]{Dimitrios Zarouchas}
\author[1]{Dimitrios Chronopoulos}

\address[1]{Department of Mechanical Engineering $\&$ Division of Mechatronic System Dynamics (LMSD), KU Leuven, Belgium}

\address[2]{Center of Excellence in Artificial Intelligence for Structures, Prognostics \& Health Management, Aerospace Engineering Faculty, Delft University of Technology, Kluyverweg 1, Delft, 2629 HS, the Netherlands}

\address[3]{Department of Civil Engineering, The University of British Columbia, Vancouver, Canada}

\address[4]{State Key Laboratory of Structural Analysis, Optimization and CAE Software for Industrial Equipment, Department of Engineering Mechanics, Dalian University of Technology, Dalian 116023, China}

\address[5]{Global College, Shanghai Jiao Tong University, Shanghai, 200240, China}

\cortext[correspondingauthor]{Corresponding author}

\begin{abstract}
Structural health monitoring (SHM) has emerged as an essential tool for ensuring the integrity and reliability of critical engineering structures, particularly in aerospace applications. Since each sensing technology has its limitations, the fusion of different modalities enables capturing a more complete picture of inhomogeneous materials, like composites. However, effective multisensor data fusion in SHM is often hindered by heterogeneous sensing modalities that operate at disparate sampling frequencies and acquisition intervals. To address these challenges, this paper proposes a Transformer-based data fusion framework that integrates multisensor data streams from piezoelectric transducer (PZT) capturing ultrasonic guided wave signals and fiber Bragg grating (FBG) sensors for strain measurements. By incorporating an attention-mechanism visualization, the proposed framework enables transparent, multitask learning for both health indicator (HI) prediction and damage localization. The framework was experimentally validated using aircraft composite structures subjected to compression-compression fatigue cyclic loading. For HI prediction, the framework consistently achieved a mean absolute error (MAE) and root mean squared error (RMSE) below 0.1, representing a nearly 60\% performance improvement over single-sensor approaches (PZT or FBG alone) and baseline deep learning models. For damage localization, the model demonstrated the highest accuracy, maintaining an MAE and RMSE below 0.0465 and 0.1571, respectively. These results demonstrate that the proposed Transformer-based data fusion framework significantly outperforms single-source models and state-of-the-art deep learning models in both HI prediction and damage localization accuracy.

\end{abstract}


\end{frontmatter}


\section{Introduction}
\label{Sec:1}
With the increasing use of carbon-fibre reinforced polymer (CFRP) composite structures in the aerospace industry (such as the Airbus A350 and Boeing 787, which comprise $53\%$ and $50\%$ CFRP, respectively \cite{broer2022need}.), understanding their complex mechanical behavior arising from anisotropic and heterogeneous properties has become critically important \cite{aggelis2012acoustic, de2008health}. Composite materials for aircraft structures offer high stiffness, strength, fracture toughness, fatigue endurance, and corrosion resistance. Aerospace materials must withstand structural and aerodynamic loads while remaining cost-effective and manufacturable. In addition, they must also exhibit sufficient damage tolerance and durability throughout the aircraft design life, which for large commercial airliners typically ranges from 30,000 to 60,000 flight hours (approximately 25–30 years) \cite{mouritz2012materials}. During this period, the aircraft structures are expected to resist cracking, corrosion, oxidation and other forms of degradation while operating under adverse conditions that involve high mechanical loads, extreme atmospheric conditions (both freezing and high-temperature), lightning strikes, hailstorms, and exposure to potentially corrosive fluids such as jet fuel and lubricants.

These conditions gradually degrade their mechanical and functional characteristics, leading to the initiation and progression of damage, and consequently reducing the overall service life of the structure \cite{mardanshahi2025sensing}. Traditionally, preventive or scheduled maintenance strategies have been employed to mitigate catastrophic failures. However, such approaches are based on periodic inspections and even component replacement, often resulting in unnecessary downtime and increased maintenance costs. To address these limitations, the maintenance paradigm has shifted towards condition-based maintenance, wherein maintenance decisions are driven by the actual health state of the structure. In this context, structural health monitoring (SHM) has emerged as a required technology. SHM involves a comprehensive collection of techniques designed to continuously evaluate structural integrity, detect and locate damage, and provide actionable insights to support maintenance decisions, thus improving safety, reliability, and lifecycle performance \cite{YANG2026114277, yang2024transfer}. Various SHM sensing techniques exist, including ultrasonic-based \cite{fang2025using, lu2024deep, FANG2026113827}, vibration-based \cite{yang2024decision}, strain-based \cite{SOMAN2025115935}, acoustic emission \cite{du2025acoustic} and eddy current testing \cite{stoll2021embedding} approaches.

The aforementioned SHM techniques often involve heterogeneous data from diverse modalities (e.g., electrical vs. optical) and units (e.g., voltage vs. wavelength). For instance, ultrasonic testing using piezoelectric transducers (PZTs) is favored for its high sensitivity to discontinuities present in structures during its propagation  \cite{mardanshahi2025sensing}.  When the excitation wave is incident on the damaged structure, it generates transmission and reflection at the damaged position. These generated waves propagate as high-frequency guided waves in the monitored structure and can then be digitally captured using piezoelectric sensors. Based on acquired signals, various ultrasonic features can be extracted for damage assessment, including time of flight (ToF) \cite{hoseini2012estimating} , time reversal \cite{mori2019damage}, and phased-array beamforming \cite{yu2016guided}. As for another popular SHM technique, optical fiber sensors are usually operated by transmitting light through optical fibers. Their dielectric nature provides resistance to harsh conditions like corrosive environments and high temperatures, making them suitable for remote sensing applications in high-voltage situations. Fiber optic strain sensors can be categorized as single-point, quasi-distributed, or fully distributed sensors, with each type addressing specific measurement needs \cite{yassin2024fiber}. Among them, fiber Bragg grating (FBG) sensors offer quasi-distributed remote monitoring and excellent chemical stability \cite{alias2024optical, ye2014structural}. The FBG sensor can be embedded directly into the composites in the manufacturing process or can also be bonded to the surface of the composites \cite{ruzicka2016properties}. While several studies \cite{sivananth2022damage, du2022development} have utilized FBG sensors for HI prediction in aerospace structures, no existing research to date has combined PZT and FBG sensors for SHM tasks in aircraft composite structures. 

As for multi-sensor data fusion approaches, most current research still relies on a single sensing modality for damage assessment. Since different sensing modalities can carry various damage-related information, an intuitive idea is to therefore fuse these sensor modalities, so that anomalies and defects can be detected with greater accuracy, enhancing the effectiveness and reliability of SHM systems \cite{broer2022need}. To this end, various data fusion approaches have been developed for SHM systems. In particular, data-driven methods have seen significant advancement, including the use of convolutional neural networks (CNNs) \cite{HUANG2024106076}, long short-term memory (LSTM) \cite{hakimi2025deep}, and gated recurrent units (GRUs) \cite{li2025transforming}, and their variants. For instance, Liu et al.~\cite{ZHENG2026122376} developed a hybrid network of the generative adversarial network (GAN) and CNN to detect and classify sensor anomalies. Also, there are works \cite{nong2023multimodal} proposed to combine 1D and 2D sensor data based on CNN. However, current literature lacks an end-to-end method capable of fusing these disparate sensor sources while simultaneously addressing the unique acquisition profiles inherent in fatigue loading scenarios.

To address these challenges, this paper proposes a Transformer-based data fusion framework that combines PZT and FBG-based sensor measurements. This framework enables end-to-end multitask learning, specifically simultaneous health indicator (HI) prediction and damage localization within SHM systems. The main contributions of this work are as follows:
\begin{itemize}
    \item A Transformer-based end-to-end data fusion framework designed to integrate PZT and FBG sensor measurements while explicitly accounting for differences in their acquisition intervals during the fatigue loading process.

    \item A multitask learning architecture that enables the SHM system to perform simultaneous HI prediction and damage localization within a unified framework.

    \item An interpretable analysis via the Transformer attention module that visualizes the interactions between PZT and FBG modalities to demonstrate the effectiveness of the fusion process.

    \item Comprehensive evaluation and benchmarking of the proposed framework against single-modality configurations as well as state-of-the-art DNN models.
    
\end{itemize}

The remainder of this work is organized as follows. Section \ref{Sec:2} details the architecture of the proposed Transformer-based framework, including the tokenization module, cycle-aware temporal encoding, and the multitask learning loss function. Section \ref{Sec:3} describes the experimental setup. Section \ref{Sec:4} validates the framework using experimental datasets and introduces quantitative metrics for HI prediction and damage localization. Finally, Section \ref{Sec:5} provides concluding remarks.

\section{Transformer-based multi-sensor data fusion framework}
\label{Sec:2}

\subsection{Problem formulation}
For data fusion within a Transformer-based architecture, we consider a multisensory SHM system where the input data may contain incomplete modalities, such as missing sensors or data acquisition occurring at different fatigue cycle intervals. Let the input data $\mathbf{X}_i$ consist of two distinct modalities: PZT sensing and FBG sensing. Because these sensing methods often operate at different intervals, we introduce a zero-patching strategy to account for scenarios where a specific modality is unavailable. The input for the $i$-th measurement is defined as:
\begin{equation}
    \mathbf{X}_i = \Big\{\mathbf{x}_{i}^{\text{pzt}} , \;\mathbf{x}_i^{\text{fbg}} \Big\}
\end{equation}
where $\mathbf{x}_{i}^{\text{pzt}}$ represents the PZT data and $\mathbf{x}_i^{\text{fbg}}$ represents the FBG strain measurements. If a modality is missing, its corresponding tensor is replaced by a zero-like tensor of identical shape to maintain architectural consistency. The objective is to map the fused modalities through a Transformer network $\mathcal{F}_{\theta}$ to perform HI prediction and damage localization tasks simultaneously:
\begin{equation}
    \mathcal{F}_{\theta}: \mathbf{X}_i \rightarrow \Big\{\mathbf{Y}_{i}^{\mathrm{det}}, \;\mathbf{G}_{i}^{\mathrm{loc}}\Big\}
\end{equation} 
which jointly optimizes for the following downstream tasks:
\begin{itemize}
    \item \textbf{HI prediction} $\mathbf{Y}_{i}^{\mathrm{det}}$: prediction of the HI progression;
    \item \textbf{Damage localization} $\mathbf{G}_i^{\mathrm{loc}} \in [0, 1]^{h \times w}$: prediction of a pixel-wise damage localization heatmap.
\end{itemize}

\subsection{Transformer-based multi-sensor data fusion framework}
\begin{figure}[H]
    \centering
    \makebox[\textwidth][c]{
    \includegraphics[width=1.2\linewidth]{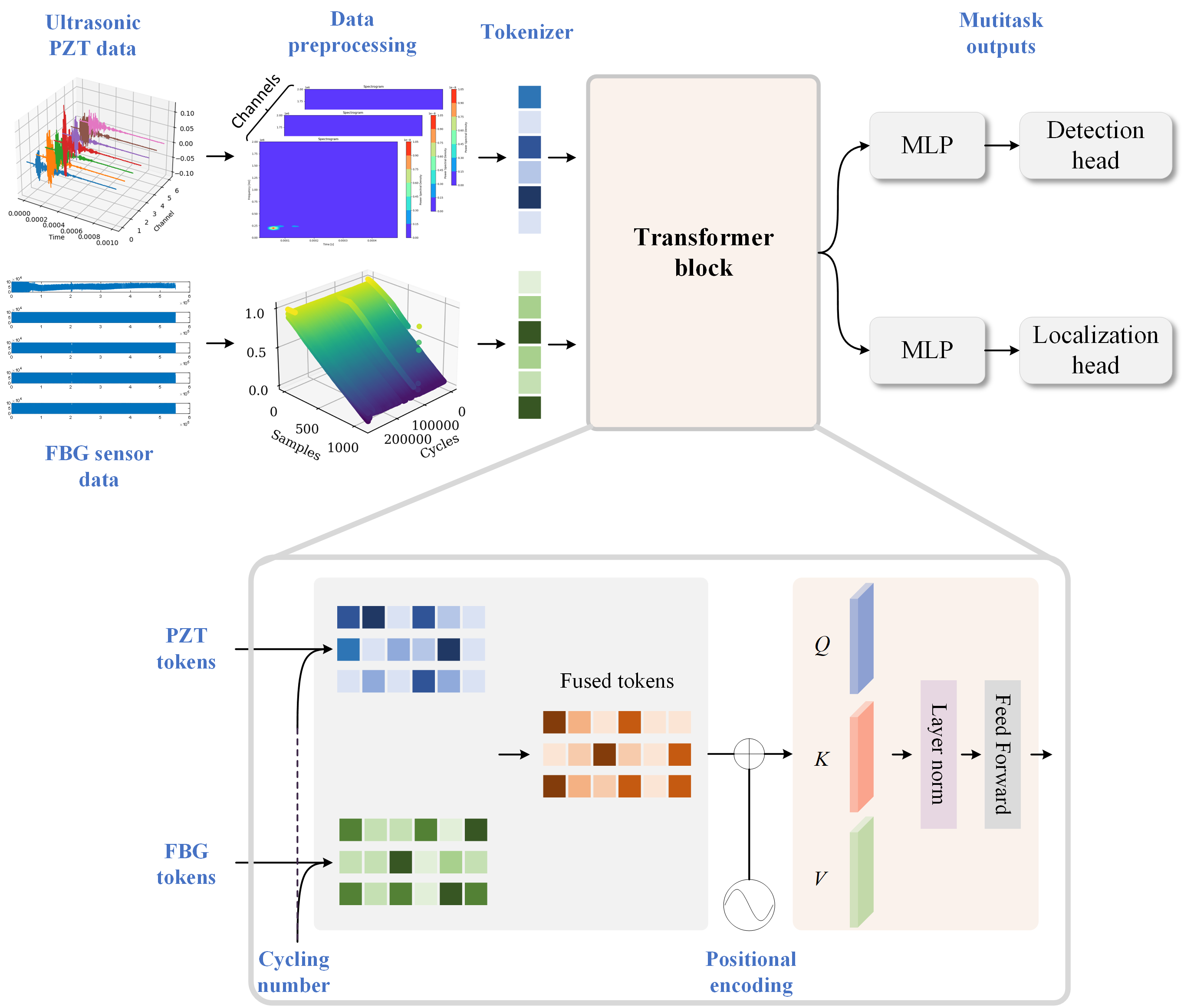}}
    \caption{\textbf{Multisensor data fusion based on the Transformer architecture.} PZT and FBG sensor data are tokenized and fused within a Transformer network, accounting for their differing acquisition intervals under specific fatigue cycles. The model then performs two tasks: HI prediction and damage localization.}
    \label{fig:data-level fusion}
\end{figure}

The proposed framework as shown in Fig.~\ref{fig:data-level fusion} relies on a unified tokenization strategy followed by a specific masking mechanism to fuse the information from the PZT and FBG sensing modalities. The architecture consists of three main stages: masked token fusion, Transformer-based data fusion, and multi-task prediction.

\subsubsection{Multisensor data tokenization}
The framework utilizes a unified tokenization and padding mechanism to transform heterogeneous raw signals into a fixed-length latent sequence suitable for Transformer processing. This process ensures data synchronization and alignment, as illustrated in the lower part of Fig.~\ref{fig:data-level fusion}. The raw signals are transformed into sequence tokens. The UGW data $\mathbf{x}_i^{\text{pzt}}$ acquired from PZT sensors are processed via short-time Fourier transform (STFT) and flattened into a sequence of $L_u$ tokens denoted $\mathbf{Z}_{\text{pzt}}$:
\begin{equation}
    \mathbf{Z}_{\text{pzt}} = [z_{\text{pzt}}^1, z_{\text{pzt}}^2, ..., z_{\text{pzt}}^{L_u}] 
\end{equation}
where $\mathbf{Z}_{\text{pzt}} \in \mathbb{R}^{L_u \times d}$ denotes the PZT tokens and $L_u$ is the number of tokens extracted from the STFT spectrogram patching. Similarly, Strain data from FBG sensors are preprocessed (in the format of 10 channel time series), segmented, and linearly projected to the same latent dimension $d$ to produce the token sequence:
\begin{equation}
    \mathbf{Z}_{\text{fbg}} = [z_{\text{fbg}}^1, z_{\text{fbg}}^2, ..., z_{\text{fbg}}^{L_s}] 
\end{equation}
where $\mathbf{Z}_{\text{fbg}} \in \mathbb{R}^{L_s \times d}$ denotes the FBG tokens. The final raw input sequence to the Transformer $\mathbf{Z}_{raw}^{(i)}$ is obtained by concatenating the tokens from both modalities to represent the fusion at the early stage:
\begin{equation}
    \mathbf{Z}_{raw}^{(i)} = \text{Concat}(\mathbf{Z}_{\text{pzt}}, \mathbf{Z}_{{\text{fbg}}}) \in \mathbb{R}^{L_{in} \times d}
\end{equation}
where $L_{in}$ is the total count of valid tokens for sample $i$. If a modality is missing, its corresponding positions are filled with zero padding. To satisfy the batch processing requirements of the Transformer, we define a fixed maximum sequence length $T_{max}$. The final input sequence $\mathbf{Z}_{in}^{(i)}$ is constructed by truncating or padding $\mathbf{Z}_{raw}^{(i)}$:
\begin{equation}
    \mathbf{Z}_{in}^{(i)} = \begin{cases} 
        \mathbf{Z}_{raw}^{(i)}\Big[1:T_{max}\Big] & \text{if } L_i > T_{max} \\
        \Big[\mathbf{Z}_{raw}^{(i)} ;\; \mathbf{P}_{pad}\Big] & \text{if } L_i \le T_{max}
        \end{cases}
\end{equation}
where $\mathbf{P}_{pad} \in \mathbb{R}^{(T_{max} - L_i) \times d}$ represents the padding tokens. Simultaneously, a binary mask $\mathbf{m}^{(i)} \in \{0, 1\}^{T_{max}}$ is generated to differentiate valid data from padding. This mask is crucial for two purposes: i) to align the attention of the Transformer to valid tokens, and ii) to normalize the loss contribution. The mask element at position $t$ is defined as:
\begin{equation}
    m_{t}^{(i)} = \mathbb{I}(t \le L_i) = \begin{cases} 
        1 & \text{if token at } t \text{ is valid (PZT or FBG)} \\
        0 & \text{if token at } t \text{ is padding}
    \end{cases}
\end{equation}
where $\mathbb{I}$ is the indicator function.

\subsubsection{Cycle-aware temporal encoding}
Once aligned, the raw tokens are transformed into rich embeddings that integrate modality-specific and temporal cycle information. As shown in Fig.~\ref{fig:data-level fusion}, the input sequence $\mathbf{Z}_{in}$ is structured as:
\begin{equation}
\begin{split}
    \mathbf{Z}_{in} &= \Big[ \underbrace{z_{\text{pzt}}^1, \dots, z_{\text{pzt}}^{L_u}}_{\text{PZT tokens}} , \underbrace{z_{\text{fbg}}^1, \dots, z_{\text{fbg}}^{L_s}}_{\text{FBG tokens}} \Big] \\
    &= [\mathbf{v}_1, \mathbf{v}_2, \dots, \mathbf{v}_{T_{max}}]
\end{split}
\label{Eq_cont_vec}
\end{equation}
To prepare these raw vectors for the Transformer, each token $\mathbf{v}_t$ is projected into the model's latent space. The initial hidden state $\mathbf{h}_t^{(0)}$ is computed as:
\begin{equation}
    \mathbf{h}_t^{(0)} = \text{LayerNorm}\Big( \mathbf{W}_c \mathbf{v}_t + \mathbf{emb}_{time}(\tau_t) + \mathbf{emb}_{modality}(m_t) \Big)
\end{equation}
where $\mathbf{W}_c \in \mathbb{R}^{d \times d_{raw}}$ is the linear projection matrix. The term $\mathbf{emb}_{time}(\tau_t)$ represents a continuous embedding of the fatigue cycle $\tau_t$, while $\mathbf{emb}_{modality}(m_t)$ distinguishes between sensor types ($m_t \in \{\text{PZT}, \text{FBG}, \text{PAD}\}$). The individual hidden states are then packed into the initial input matrix $\mathbf{H}^{(0)} = [\mathbf{h}_1^{(0)}, \dots, \mathbf{h}_{T_{max}}^{(0)}]^{\top} \in \mathbb{R}^{T_{max} \times d}$, which serves as the entry for the first Transformer layer.

\subsubsection{Cross-modal attention module}
The core of the framework is the Transformer framework, which performs cross-modal fusion through a multi-head self-attention (MHSA) mechanism \cite{petit2021u}. As illustrated in the expanded view of Fig.~\ref{fig:data-level fusion}, the fused sequence comprising PZT and FBG tokens is combined with positional encoding. For each attention head, the input embeddings are linearly projected into three distinct matrices: Queries ($\mathbf{Q}$), Keys ($\mathbf{K}$), and Values ($\mathbf{V}$)
\begin{equation}
    \mathbf{Q} = \mathbf{H}^{(l-1)}\mathbf{W}_Q, \quad \mathbf{K} = \mathbf{H}^{(l-1)}\mathbf{W}_K, \quad \mathbf{V} = \mathbf{H}^{(l-1)}\mathbf{W}_V
\end{equation}
where $\mathbf{W}_Q, \mathbf{W}_K, \mathbf{W}_V \in \mathbb{R}^{d \times d_k}$ are the learnable projection weights. The attention mechanism computes the compatibility between the PZT and FBG modalities by calculating the dot product of the queries and keys. To prevent vanishing gradients during training, the scores are scaled by $\sqrt{d_k}$ \cite{vaswani2017attention, yang2025damage}:
\begin{equation}
    \text{Attention}(\mathbf{Q}, \mathbf{K}, \mathbf{V}) = \text{Softmax}\left(\frac{\mathbf{QK}^{\top}}{\sqrt{d_k}}\right)\mathbf{V}
\end{equation}
where $d_k$ refers to the dimension of the key matrix $\mathbf{K}$. The MHSA computes the attention matrix $\mathbf{QK}^{\top}$, which calculates a similarity score between all tokens:
\begin{itemize}
    \item \textbf{Intra-modal} (PZT $\leftrightarrow$ PZT and FBG $\leftrightarrow$ FBG): Captures localized dependencies and signal patterns within each independent sensor modality.
    \item \textbf{Cross-modal} (PZT $\leftrightarrow$ FBG and FBG $\leftrightarrow$ PZT): Acts as the fusion engine by mapping nonlinear correlations between the ultrasonic spectral features and FBG temporal events.
\end{itemize}

\subsubsection{Multi-task learning loss design}
To learn multitask HI prediction and damage localization, we define a unified multi-task loss $\mathcal{L}_{tot}$ to train the model. This loss leverages the binary mask $\mathbf{m}^{(i)}$ generated during the tokenization process to ensure that gradients are only calculated from valid measurements (because valid measurements from PZT and FBG sensors are very sparse), thereby preventing corruption from padded or missing data. The total loss is a weighted linear combination of the three main loss functions:
\begin{equation}
    \mathcal{L}_{tot} = \lambda_{det}\mathcal{L}_{det} + \lambda_{loc}\mathcal{L}_{loc} + \lambda_{rec}\mathcal{L}_{rec}
\label{Eq4-9}
\end{equation}
where $\lambda_{det}$, $\lambda_{loc}$ and $\lambda_{rec}$ represent the weighting parameters for HI prediction, localization and content reconstruction loss, respectively. The content reconstruction loss ($\mathcal{L}_{rec}$) serves as a self-supervised regularizer to counter representation collapse resulting from highly sparse and asymmetrical PZT and FBG tokens. By penalizing the content vector $\mathbf{v}_i$ in Eq.~\ref{Eq_cont_vec}, the embedding space can preserve high-fidelity physical characteristics of the raw sensor modalities, a strategy used in multisensor data fusion to significantly reduce modality gaps and improve robustness against missing or non-aligned sensor sequences \cite{woo2023goodpracticesmissingmodality, wu2026deepmultimodallearningmissing}.

Specifically, the mask $\mathbf{m}^{(i)}$ is applied to normalize the contribution of each sample $i$, effectively calculating the loss over the effective batch size $N_{eff}$:
\begin{equation}
    \mathcal{L}_{tot} 
             = \frac{1}{N_{eff}} \sum_{i=1}^{N} \mathbb{I}(\mathbf{m}^{(i)}) \left[\lambda_{det}\mathcal{L}_{det}(\mathbf{Y}^{\text{det}}_i, \hat{\mathbf{Y}}^{\text{det}}_i) + \lambda_{loc}\mathcal{L}_{loc}(\mathbf{G}^{\text{loc}}_i, \hat{\mathbf{G}}^{\text{loc}}_i) +
             \lambda_{rec}\mathcal{L}_{rec}(\mathbf{v}_i, \hat{\mathbf{v}}_i) \right]
\end{equation}
where $\mathbf{Y}_i^{\text{det}}$ and $\hat{\mathbf{Y}}_i^{\text{det}}$ denote the ground truth and predicted HI values for HI prediction, respectively. Similarly, $\mathbf{G}_i$ and $\hat{\mathbf{G}}_i$ represent the ground truth and predicted damage localization maps.

\section{Experiment setup}
\label{Sec:3}
As part of the ReMAP Horizon 2020 EU project, multiple rounds of testing were conducted at the TU Delft Aerospace Structures and Materials Laboratory and Department of Mechanical Engineering \& Aeronautics of University of Patras, respectively \cite{Agnes2022, GALANOPOULOS2023116579, s21175701}. Overall, the database consists of multiple SHM sensing data that were collected from the composite skin-stiffener panels going through compression-compression (C-C) fatigue loading tests. 

\begin{figure}[H]
    \centering
    \makebox[\textwidth][c]{
    \includegraphics[width=1.1\linewidth]{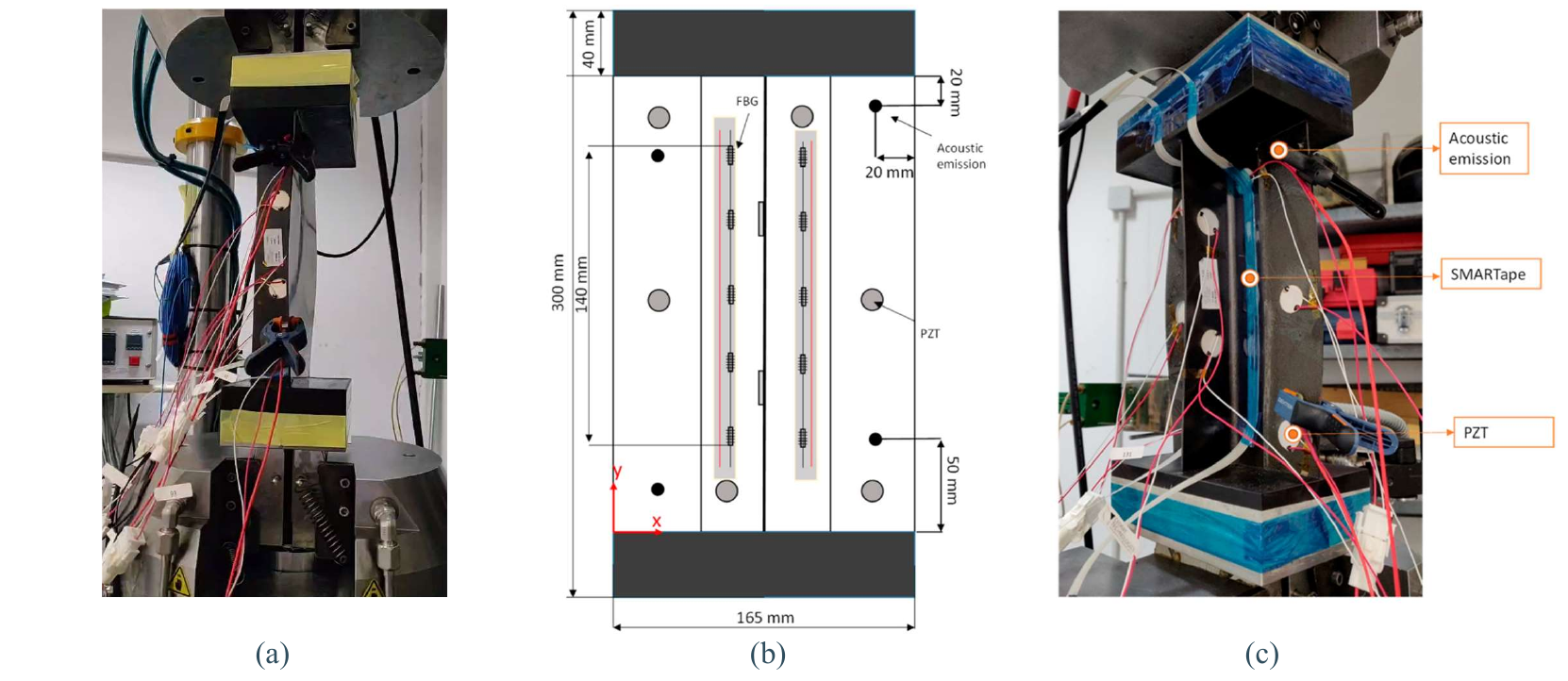}}
    \caption{The experimental setup in the ReMAP dataset \cite{Agnes2022, GALANOPOULOS2023116579, s21175701}. (a) ReMAP datasets experiment campaign; (b) An illustration of the geometry and sensor locations of tested composite panels; (c) Visualization of sensor location in experiment.}
    \label{fig:ReMAP}
\end{figure}

The coupons were tested until failure, and these composite panels with Embraer designs consist of a single-layer skin panel with T-shaped stiffeners. Both skin and stiffener were fabricated using $\mathrm{IM7/8552}$ carbon fiber-reinforced epoxy unidirectional prepreg with layup information specified in Table~\ref{tab:ReMap_layup}. Each composite specimen incorporated two resin blocks to ensure a uniform load distribution. The dimensions of a single panel are shown in Fig.~\ref{fig:ReMAP}(b). All panels were subjected to a 10 $\mathrm{J}$ impact load applied at varying locations and time points in different regions of the skin and the stiffener. 

\begin{figure}[H]
    \centering
    \makebox[\textwidth][c]{
    \includegraphics[width=1\linewidth]{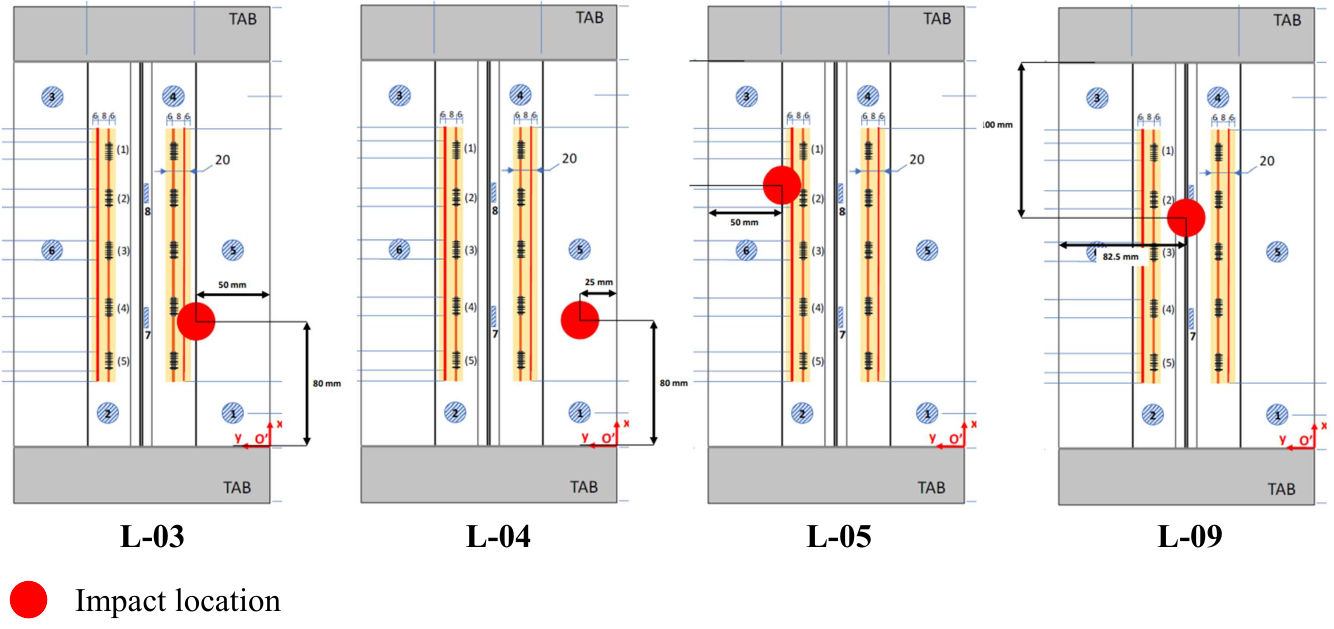}}
    \caption{The initial impact locations for the ReMAP composite panels \cite{Agnes2022, GALANOPOULOS2023116579, s21175701}. Six PZT sensors were evenly distributed on the surface of the composite panels, and two FBG each integrated with five strain gauges were located at the left and right foot of the stiffener. A 10 $\mathrm{J}$ impact load was applied at varying locations denoted with red round shape.}
    \label{fig:ReMAP_specimens}
\end{figure}

\begin{table}[H]
\centering
\renewcommand{\arraystretch}{2}
\scriptsize
\caption{Layup information and material property in ReMAP composite panels}
\begin{tabularx}{\linewidth}{@{} >{\centering\arraybackslash}m{3cm} @{}>{\centering\arraybackslash}m{4.5cm} @{}>{\centering\arraybackslash}m{4.5cm} @{}>{\centering\arraybackslash}m{5cm} @{}}
\toprule
Specimens & Skin layups & Stiffener layups & Materials \\
\midrule

L03
  & \multirow{4}{*}{$[45/-45/0/45/90/-45/0]_s$}
  & \multirow{4}{*}{$[45/-45/0/45/-45]_s$}
  & \multirow{4}{5cm}{\centering $\mathrm{IM7/852}$\\carbon fiber-reinforced\\unidirectional prepreg} \\

L04 & & & \\

L05 & & & \\

L09 & & & \\
\bottomrule
\end{tabularx}
\label{tab:ReMap_layup}
\end{table}

The ReMAP project employs six distinct SHM sensor technologies. This study focuses on the following three sensing technologies for tracking the progression of internal damage within composite panels: 
\begin{itemize}
    \item Lamb wave detection system (LWDS);
    \item Fiber Bragg gratings (FBGs);
    \item Acoustic emission (AE);
\end{itemize}
In the fatigue testing, PZT sensors employed within the LWDS and FBG sensors serve as input sensor sources, while AE characteristics throughout the fatigue loading process function as labels. It is particularly noted that PZT and FBG data were collected after every 5000 and 500 fatigue loading cycles, respectively. When the LWDS captured PZT signals, the FBG sensors were inactive \cite{yue2022assessing}. In the experiment, two smart tapes, each containing five FBGs, were affixed to the left and right sides of the stiffener. Sensor information and specific locations are detailed in Fig.~\ref{fig:ReMAP}(b-c) and Table~\ref{tab:ReMAP_Sen}.
\begin{table}[H]
\renewcommand{\arraystretch}{2}
\scriptsize
\caption{Sensor information in ReMAP dataset}
\noindent\makebox[\linewidth][c]{%
  \resizebox{1\linewidth}{!}{%
    \begin{tabularx}{1.15\linewidth}{@{\extracolsep{\fill}} c c c c c c c c c c}
        \toprule
        \multicolumn{2}{c}{Sensor types} & \multicolumn{2}{c}{Interval} &
        \multicolumn{2}{c}{Time of measure} & \multicolumn{2}{c}{Measuring rate} &
        \multicolumn{2}{c}{I/O} \\
        \midrule
        \multicolumn{2}{c}{PZT} & \multicolumn{2}{c}{
        every 5000 cycles} &
        \multicolumn{2}{c}{At 0 mm} & \multicolumn{2}{c}{N/A} &
        \multicolumn{2}{c}{Input} \\
        \multicolumn{2}{c}{FBG} & \multicolumn{2}{c}{
        every 500 cycles} &
        \multicolumn{2}{c}{Continuous during quasi-static} & \multicolumn{2}{c}{100 Hz} &
        \multicolumn{2}{c}{Input} \\
        \multicolumn{2}{c}{AE} & \multicolumn{2}{c}{
        Continuous} &
        \multicolumn{2}{c}{N/A} & \multicolumn{2}{c}{N/A} &
        \multicolumn{2}{c}{Output} \\
        \bottomrule
    \end{tabularx}
  }
}
\label{tab:ReMAP_Sen}
\end{table}

\subsection{AE-driven labeling for health indicator}
As mentioned previously, AE sensors were constantly operating throughout the entire fatigue tests. Therefore, it is reasonable to extract AE-based features to construct HI that can track damage evolution. Since the purpose of an HI is to reflect the progression of structural degradation, selecting an appropriate AE feature is essential for reliable downstream learning in the Transformer model. Recent works \cite{MORADI2023105502, MORADI2025112156, moradi2024novel} have suggested three criteria for evaluating the suitability of an AE-based HI: monotonicity (Mo), prognosability (Pr), and trendability (Tr).

i) \textbf{Monotonicity (Mo)}: The average monotonicity for a population of 
$N$ nominally identical structures is defined as follows \cite{MORADI2025112156}:
\begin{equation}
    \text{Mo} = \frac{1}{M} \sum_{j=1}^{M} \left| \frac{1}{N_j - 1} \sum_{i=1}^{N_j} 
    \frac{\displaystyle \sum_{p=1,\,p>i}^{N_j} (t_p - t_i) \cdot sgn\bigl( x(t_p) - x(t_i) \bigr)}
         {\displaystyle \sum_{p=1,\,p>i}^{N_j} (t_p - t_i)} \right|
\end{equation}
where $x(t_p)$ and $x(t_i)$ denote the measurements at the times of $t_p$ and $t_i$, respectively. The $sgn(\cdot)$ represents the signum function.

ii) \textbf{Prognosability (Pr)}: Prognosability is defined as the standard deviation of the HI failure values of the available lifetimes, divided by the average variation between the values of the HI at the start and end of the lifetime (failure value). To ensure the scale is between $[0, 1]$, the metric is exponentially weighted \cite{GALANOPOULOS2023116579}.

\begin{equation}
    \text{Pr} = \exp\left(
    - \frac{
            \sqrt{
                \frac{1}{M} \sum_{j=1}^{M}\left[ x_{j}(N_{j})
                - \frac{1}{M} \sum_{i=1}^{M} x_{i}(N_{i})\right]^{2}
            }     
    }{
        \frac{1}{M} \sum_{j=1}^{M} \left| x_{j}(1) - x_{j}(N_{j}) \right|
    }
    \right)
\end{equation}
iii) \textbf{Trendability (Tr)}: Trendability captures how similarly the HI evolves across different structures, independent of absolute scale. It is defined based on the correlation of each structure’s HI trajectory with those of all others \cite{MORADI2026112767}:
\begin{equation}
    \text{Tr} = \underset{j,k}{\mathrm{min}} \Bigg|\frac{cov(x_i, x_j)}{\sigma_{x_j}\sigma_{x_k}}\Bigg|
\end{equation}
where $cov(x_i, x_j)$ signifies covariance, where $x_j$ and $x_k$ are vectors of measurements for the $j^{th}$ and $k^{th}$ unit (out of the $M$ units) with measurements $N_j$ and $N_k$, respectively. The standard deviations of $x_j$ and $x_k$ are separately denoted by  $\sigma_{x_j}$ and $\sigma_{x_k}$.

Following the evaluation of multiple AE features using the Mo–Pr–Tr criteria, cumulative energy was selected as the most effective HI for fatigue loading. The cumulative energy naturally increases as damage progresses \cite{DU2024110238}, typically shows low fluctuation at failure, and exhibits similar growth patterns across multiple specimens, which lead to high monotonicity, strong trendability, and favorable prognosability.

\begin{figure}[H]
    \centering
    \makebox[\textwidth][c]{
    \includegraphics[width=1.1\linewidth]{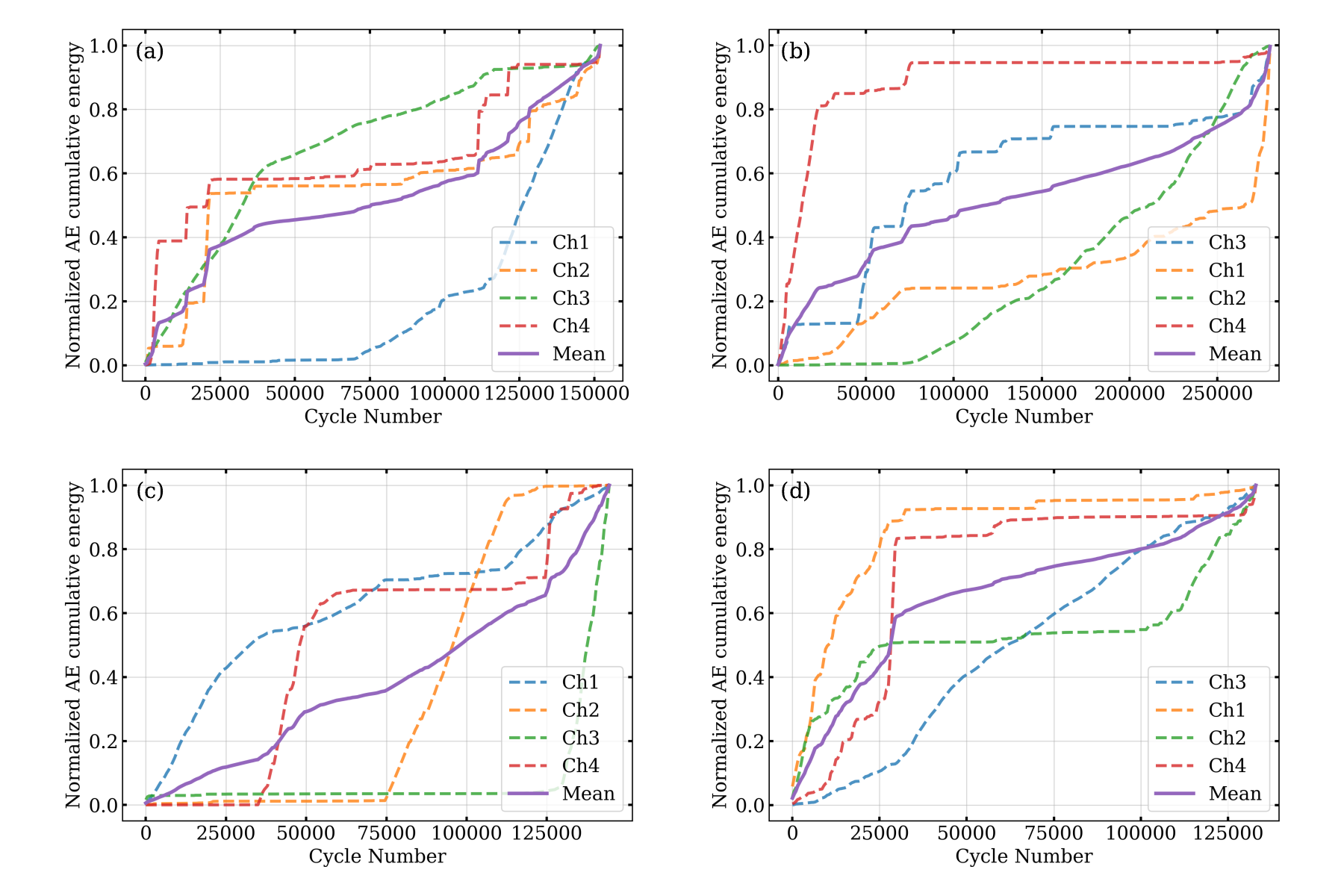}}
    \caption{Normalized AE cumulative energy considered as HI labels. Subfigures (a-d) represent the normalized AE cumulative energy for L03, L04, L05, and L09 composite specimens, respectively. The mean curve in each subfigure was obtained by averaging cumulative energy from four channels of AE sensors.}
    \label{fig:AE_HI}
\end{figure}

In this regard, the cumulative energy extracted from AE data was selected as a useful label generator for HI. Since four AE sensors were installed on the structural surface, each composite panel could collect four-channel AE data. The resulting cumulative energy curve (see Fig.~\ref{fig:AE_HI}) demonstrates an upward trend across all four AE data channels when cumulative energy is employed as a damage indicator. According to the criteria of Mo, Pr, and Tr mentioned above, the damage indicators derived from each channel were averaged to produce a mean damage curve. This curve will serve as the HI label within the subsequent Transformer network.

\subsection{AE-driven labeling for damage localization}
Besides HI prediction, AE data can be employed for damage localization. In order to determine the location of possible damage, AE sensors need to be installed at different locations on the surface of the tested composite specimens. The transient waves emitted by the acoustic emission source arrive at the sensors at different times. In the experiment campaign, the AMSY-6 software automatically groups the probe waves from the same acoustic emission source into an event set \cite{galanopoulos2023acoustic}. The arrival times of the hits in the event set lead to a series of arrival time differences. These time differences can be processed to produce a location result of the AE source.

\begin{figure}[H]
    \centering
    \makebox[\textwidth][c]{
    \includegraphics[width=1\linewidth]{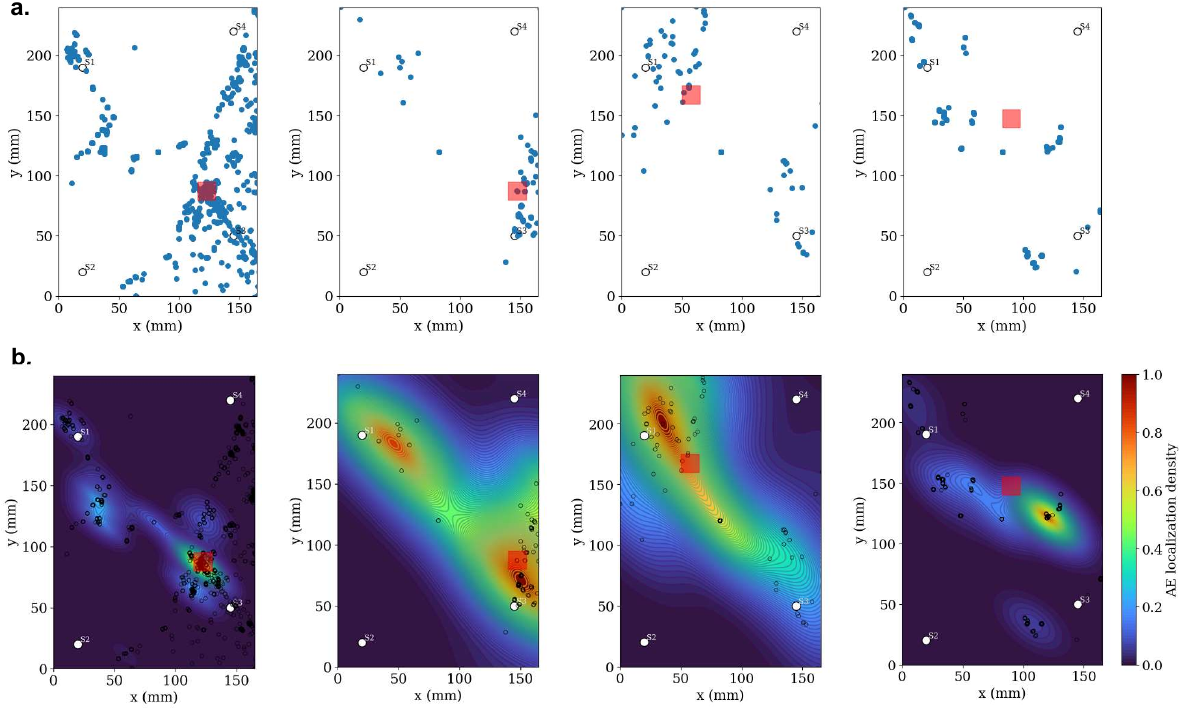}}
    \caption{Gaussian KDE-based damage localization by clustering triggered AE hits. L03, L04, L05, L09 AE localization (upper row) with corresponding Gaussian KDE clustering results (bottom row) are separately presented in a and b subfigures. The red squares denote the initial impact locations.}
    \label{fig:AE_LOC}
\end{figure}

The composite panels are orthotropic specimens with distinct wave propagation
velocities in different directions. Let the group velocity along the $x$-axis and $y$-axis be denoted by $u_x$ and $u_y$, respectively. For a damage source located at $\mathbf{\phi}=(x,y)$ and a sensor located at $\mathbf{s}_i=(x_i, y_i)$, the angle of the wave propagation path $\theta_i$ is defined as:
\begin{equation}
    \theta_i = \arctan \Big(\frac{y_i - y}{x_i - x}\Big)
\end{equation}
The effective group velocity $c(\theta_i)$ for the path between the source and the $i$-th sensor is direction-dependent and is governed by the ellipsoidal velocity profile \cite{raorane2025acoustic}:
\begin{equation}
    c(\theta_i) = \left( \frac{\cos^2\theta_i}{u_x^2} + \frac{\sin^2\theta_i}{u_y^2} \right)^{-1/2}
\end{equation}
The above equation accounts for wave velocity variations caused by material anisotropy, ensuring that ToF calculations reflect the directional dependence of composite laminates. Raw AE hits are categorized as discrete events based on temporal proximity. Let $H = \{h_1, h_2, \dots, h_N\}$ be the time-sorted sequence of hits, where each hit $h_k$ possesses a timestamp $t_k$ and a channel index $id_k$. If a consecutive subset of hits $E_m = \{h_j, \dots, h_{j+k}\}$ satisfies the temporal continuity constraint, it is defined as a candidate event window:
\begin{equation}
    t_{n+1} - t_n \leq \tau_{tol} \quad \forall \ h_n, h_{n+1} \in E_m
\end{equation}
where $\tau_{tol}$ is a pre-defined tolerance. To filter out noise and incomplete data, a candidate window is accepted as a valid event only if the number of unique sensor channels within the window satisfies $| \{id_n \mid h_n \in E_m\} | \geq N_{min}$, where $N_{min}$ is the minimum required sensors for triangulation.
Localization is performed using a Time Difference of Arrival (TDoA) approach to eliminate the unknown emission time $t_0$ \cite{ciampa2012impact, raorane2025acoustic}. For a specific subset of sensors $S_{sub} \subseteq S$, the theoretical travel time $T_i(\mathbf{\phi})$ from the source $\mathbf{\phi}$ to sensor $i$ is:
\begin{equation}
    T_i(\mathbf{\phi}) = \frac{\| \mathbf{s}_i - \mathbf{\phi} \|}{c(\theta_i)}
\end{equation}
The residual vector $\mathbf{r}(\mathbf{\phi})$ based on the difference between the predicted differential arrival times and the measured differential arrival times. Using the first sensor in the subset (index $0$) as the reference:
\begin{equation}
    r_i(\mathbf{\phi}) = \Big(T_i(\mathbf{\phi}) - T_0(\mathbf{\phi})\Big) - \Big(t_i^{meas} - t_0^{meas}\Big)
\end{equation}
The estimated source location $\hat{\mathbf{\phi}}$ is obtained by minimizing the sum of squared residuals using the Levenberg-Marquardt algorithm \cite{gavin2019levenberg}:
\begin{equation}
    \hat{\mathbf{\phi}} = \operatorname*{argmin}_{\mathbf{\phi}} \sum_{i \in S_{sub}} \Big( r_i(\mathbf{\phi}) \Big)^2
\end{equation}
For detected AE events involving a set of active sensors $S_{active}$, it can generate all possible combinations of sensor subsets of size $k$ (where $k \ge N_{min}$). Let $\mathcal{C}$ denote the set of all such combinations. For each subset $C_j \in \mathcal{C}$, a candidate location $\hat{\mathbf{\phi}}_j$ and associated residual cost $J_j = \sum r^2$ can be computed. Each solution is assigned a weighting factor $w_j$ inversely proportional to the residual error, prioritizing solutions that better align with the physical model:
\begin{equation}
    w_j = \frac{1}{J_j + \epsilon}
\end{equation}
where $\epsilon$ is a small regularization constant. The final spatial distribution of the acoustic source is visualized as a probability density function $f(\mathbf{\phi})$ estimated via Gaussian kernel density estimation (Gaussian KDE) \cite{scott2015multivariate}:
\begin{equation}
    f(\mathbf{\phi}) = \sum_{j} w_j K_{\mathrm{h}}(\mathbf{\phi} - \hat{\mathbf{\phi}}_j)
\end{equation}
where $K_{\mathrm{h}}$ is the kernel function with bandwidth $\mathrm{h}$, and $w_j$ represents weight associated with each estimated source location $\hat{\mathbf{\phi}}_j$. In this work, the Gaussian KDE framework introduced is employed to aggregate all source estimates derived from the triggered AE hits. The resulting density map highlights regions of high event concentration, thereby offering a probabilistic view of the possible damage locations. The corresponding AE localization results are illustrated in Fig.~\ref{fig:AE_LOC}.
The upper row of subplots shows the estimated damage source locations for the composite panels L03, L04, L05, and L09 derived from the extracted AE arrival times. Red rectangles indicate the known pre-damaged regions. Because each AE event is detected by a minimum of three sensors, every triggered event yields one triangulated source estimate, resulting in a cloud of location points. To provide a more interpretable visualization of the underlying source distribution, Gaussian KDE clustering was applied to these point sets. The resulting density-based cluster maps are shown in the bottom row. The Gaussian KDE-based clusters exhibit a clear concentration around the true impact regions, demonstrating that the KDE-based AE localization method effectively aggregates noisy triangulated estimates and enhances the reliability of damage localization.

\section{Results and discussions}
\label{Sec:4}
\subsection{Tokenization results}
\begin{figure}[H]
    \centering
    \makebox[\textwidth][c]{\includegraphics[width=1.2\linewidth]{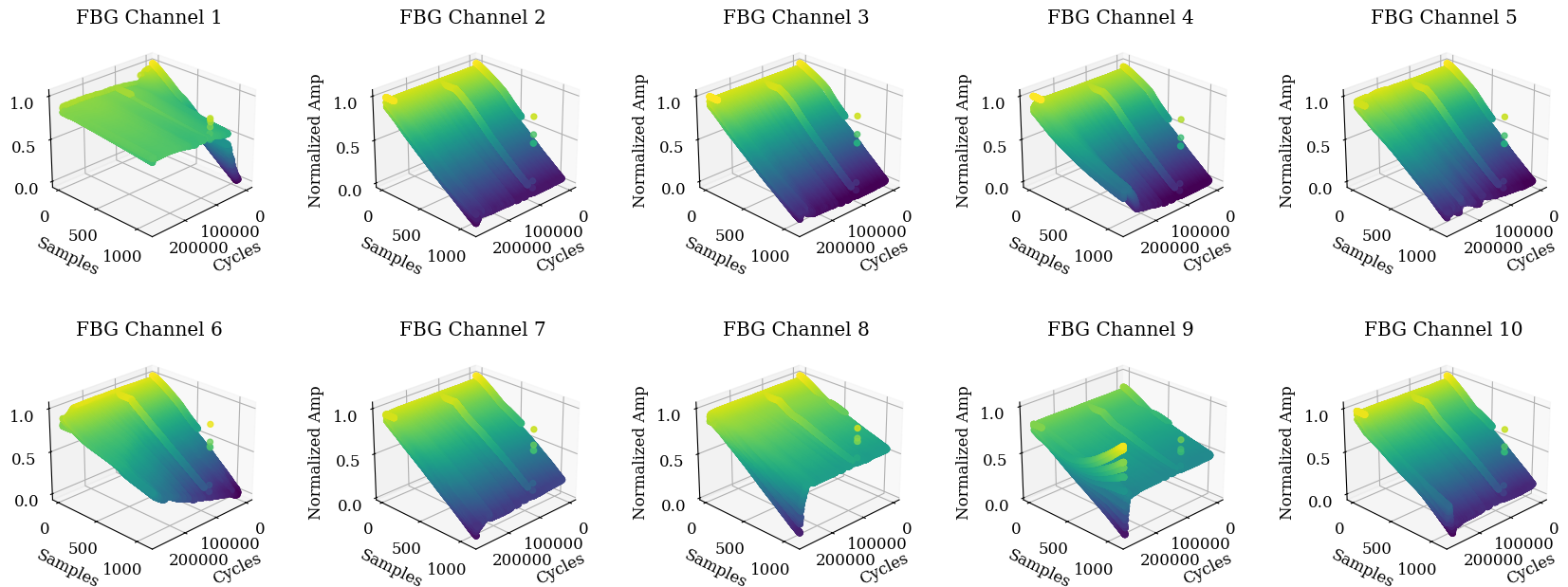}}
    \caption{10-channel FBG raw data visualization for L04 composite panel in ReMAP dataset. For each channel, there were 1100 samples at specific FBG acquisition cycle.}
    \label{fig:fbg_data}
\end{figure}

\begin{figure}[H]
    \centering
    \makebox[\textwidth][c]{
    \includegraphics[width=1\linewidth]{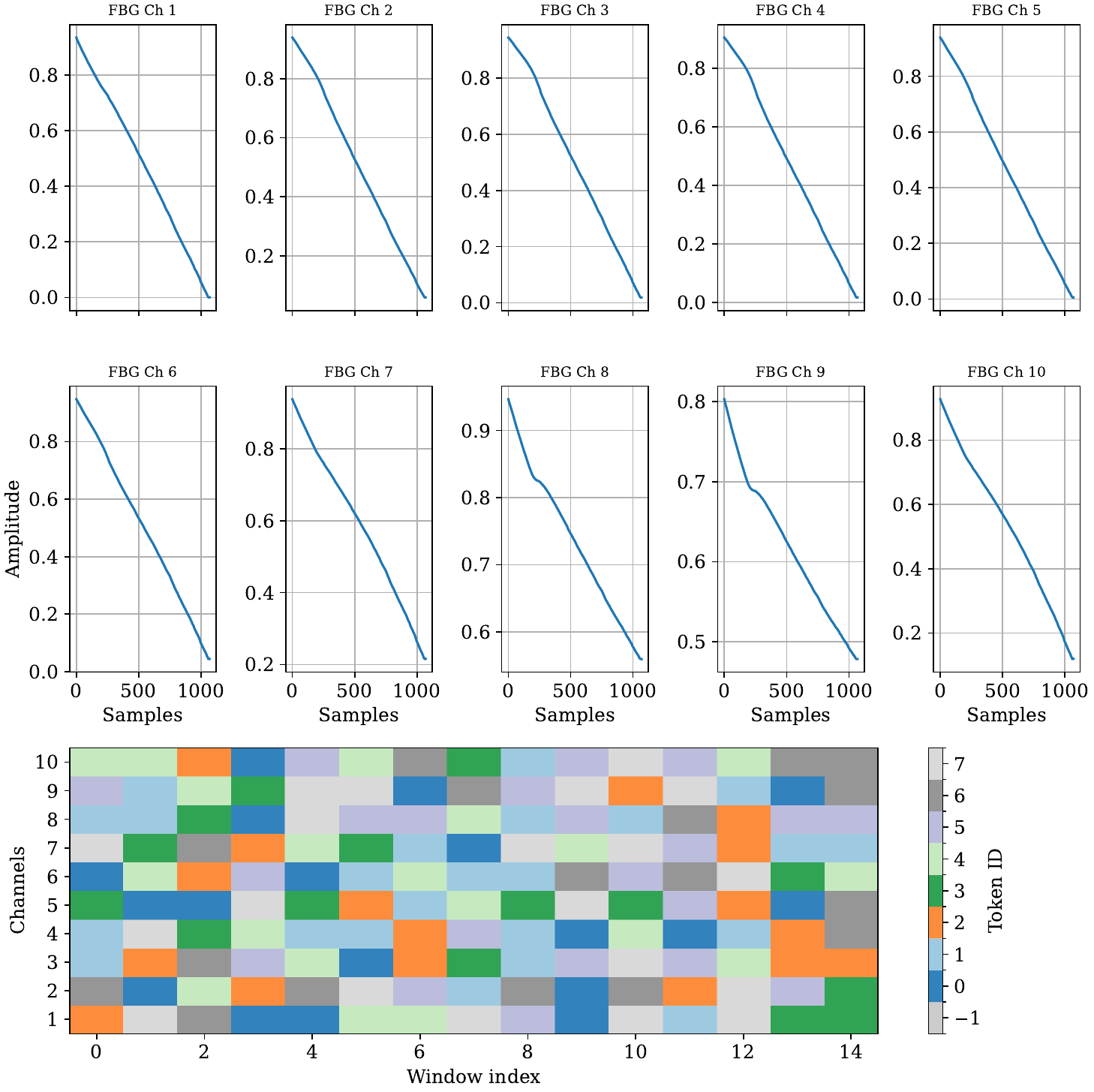}}
    \caption{The 10-channel FBG segmentation results (at 500th cycle) of the L04 composite panel in the ReMAP dataset. Considering the sampling rate, each 10-channel FBG data is visualized with 1100 samples, and the corresponding tokenization results are displayed in the bottom row.}
    \label{fig:fbg_token}
\end{figure}

Before transforming the FBG data into raw tokens usable by the Transformer network, a preprocessing step is required to separate the raw FBG data into a specific format based on its cyclic variation pattern. Since FBG data was collected during the quasi-static process where the MTS machine was proceeding with a displacement of 0.5 $\mathrm{mm/min}$, each FBG yielded a total of 1100 samples. Fig.~\ref{fig:fbg_data} visualizes the FBG raw data across 10 channels, where the x-axis represents the number of cycles and the y-axis denotes the number of sampling points. The FBG data were collected every 500 cycles and normalized to the range of 0–1. 
Fig.~\ref{fig:fbg_token} visualizes the tokenization results of 10-channel FBG data captured at the 500th cycle for the composite panel L04. Since the FBG data were acquired during a quasi-static process, each channel contains a total of 1100 data samples. In the time-series segmentation process, a sliding window strategy was adopted to generate segmented units for each channel. Consequently, the FBG data can be transformed into matrix-based tokens as shown in the bottom of the Fig.~\ref{fig:fbg_token}.

\begin{figure}[H]
    \centering
    \includegraphics[width=0.85\linewidth]{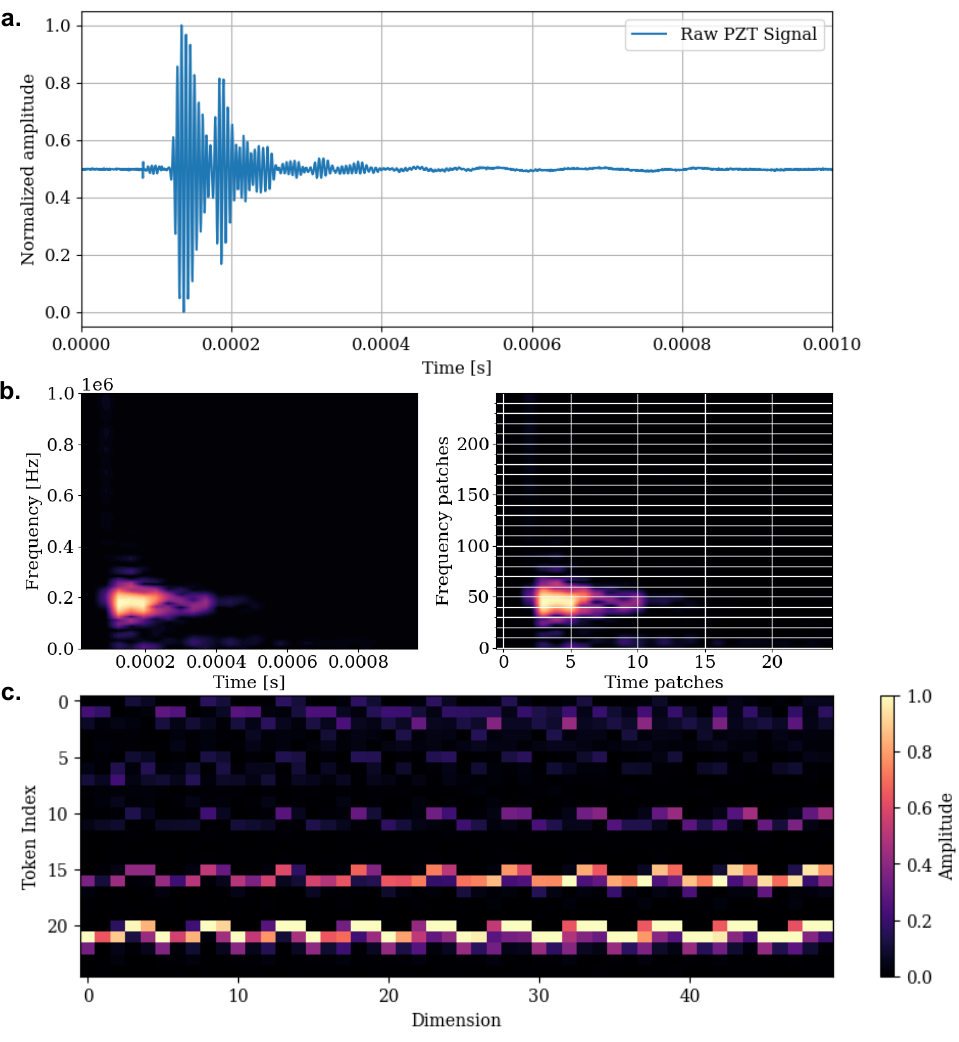}
    \caption{The multichannel UGW data tokenization results. From the raw multichannel PZT series, the STFT was implemented to transform the PZT series into a 3D time-frequency spectrum. Then the tokenization process was utilized to split the spectrum into tokens.}
    \label{fig:PZT_Tokens}
\end{figure}

Similarly, the tokenization process for the UGW series is illustrated in Fig.~\ref{fig:PZT_Tokens}. First, the raw ultrasonic guided wave signal in Fig.~\ref{fig:PZT_Tokens}(a) is projected into the time-frequency domain based on STFT, generating the spectrogram shown in Fig.~\ref{fig:PZT_Tokens}(b). This transformation converts the one-dimensional time-series data into a structured time-frequency representation. Subsequently, the tokenization process partitions this complex data into discrete, non-overlapping patches to facilitate processing by the fusion module within the Transformer framework. Fig.~\ref{fig:PZT_Tokens}(c) visualizes the first 25 resulting tokens. It can be observed that each token is flattened into a distinct 48-dimensional vector. These ultrasonic tokens are then prepared to be concatenated with tokens derived from the FBG sensors.

\subsection{Evaluation metrics}
To quantitatively assess the accuracy of HI prediction based on the AE cumulative energy–based HI, several HI prediction and damage localization performance metrics are employed. For the prediction of the HI progression, the mean absolute error (MAE), root mean squared error (RMSE), and the coefficient of determination ($\mathrm{R^2}$) are used to evaluate HI prediction accuracy. 

\begin{equation}
    \mathrm{MAE}_{\mathrm{det}} = \frac{1}{N_{\mathrm{test}}} \sum_{i=1}^{N_{\mathrm{test}}} \left| \hat{\mathbf{Y}}_{\mathrm{test},i} - \mathbf{Y}_{\mathrm{test},i} \right|
\end{equation}

\begin{equation}
\mathrm{RMSE}_{\mathrm{det}} =
    \sqrt{
    \frac{1}{N_{\mathrm{test}}} \sum_{i=1}^{N_{\mathrm{test}}} \left( \hat{\mathbf{Y}}_{\mathrm{test},i} - \mathbf{Y}_{\mathrm{test},i} \right)^2
    }
\end{equation}

\begin{equation}
\mathrm{R^2}_{\mathrm{det}} =
    1 - \frac{
    \sum_{i=1}^{N_{\mathrm{test}}} \left( \mathbf{Y}_{\mathrm{test},i} - \hat{\mathbf{Y}}_{\mathrm{test},i} \right)^2
    }{
    \sum_{i=1}^{N_{\mathrm{test}}} \left( \mathbf{Y}_{\mathrm{test},i} - \bar{\mathbf{Y}} \right)^2
    },
\quad
    \bar{\mathbf{Y}} = \frac{1}{N_{\mathrm{test}}} \sum_{i=1}^{N_{\mathrm{test}}} \mathbf{Y}_{\mathrm{test},i} 
\end{equation}
where $N_{\mathrm{test}}$ denotes test samples. $\mathbf{Y}_i$ and $\hat{\mathbf{Y}}_i$ represent ground truth and predicted values, respectively. As for damage localization, let $\mathbf{H}_i \in \mathbb{R}^{H \times W}$ and $\hat{\mathbf{H}}_i \in \mathbb{R}^{H \times W}$ denote the ground truth and predicted spatial damage maps, respectively. The localization performance is evaluated using pixel-wise mean square error (MSE) and MAE, defined as:
\begin{equation}
\mathrm{MSE}_{\mathrm{loc}} =
\frac{1}{N H W}
\sum_{i=1}^{N}
\sum_{h=1}^{H}
\sum_{w=1}^{W}
\left(
\hat{\mathbf{H}}_i^{(h,w)} - \mathbf{H}_i^{(h,w)}
\right)^2
\end{equation}

\begin{equation}
\mathrm{MAE}_{\mathrm{loc}} =
\frac{1}{N H W}
\sum_{i=1}^{N}
\sum_{h=1}^{H}
\sum_{w=1}^{W}
\left|
\hat{\mathbf{H}}_i^{(h,w)} - \mathbf{H}_i^{(h,w)}
\right|
\end{equation}
The ground truth damage localization presented is the Gaussian KDE heatmap, which can be seen as a gray-scale image. Thus, for damage localization, an image similarity metric can be introduced to further evaluate the quality of the damage localization. One known metric called the Structural Similarity Index (SSIM), which assesses the structural similarity between two images by considering luminance, contrast, and structure. It's often used to evaluate the perceived quality of an image compared to a reference image \cite{nilsson2020understandingssim}. For two images, SSIM is defined as follows \cite{1284395}.
\begin{equation}
    \mathrm{SSIM}(x, y)= \Big(l(x,y)\Big)^{\alpha} \Big(c(x,y)\Big)^{\beta} \Big(s(x,y)\Big)^{\gamma}
\end{equation}
where $l, c, s$ denote the \textit{luminance}, \textit{contrast} and \textit{structure} component of the image, and $x,y$ refer to the pixel position of the image.

\subsection{Results and comparison}
\subsubsection{Multitask training strategy}
\begin{figure}[H]
    \centering
    \makebox[\textwidth][c]{
    \includegraphics[width=0.9\linewidth]{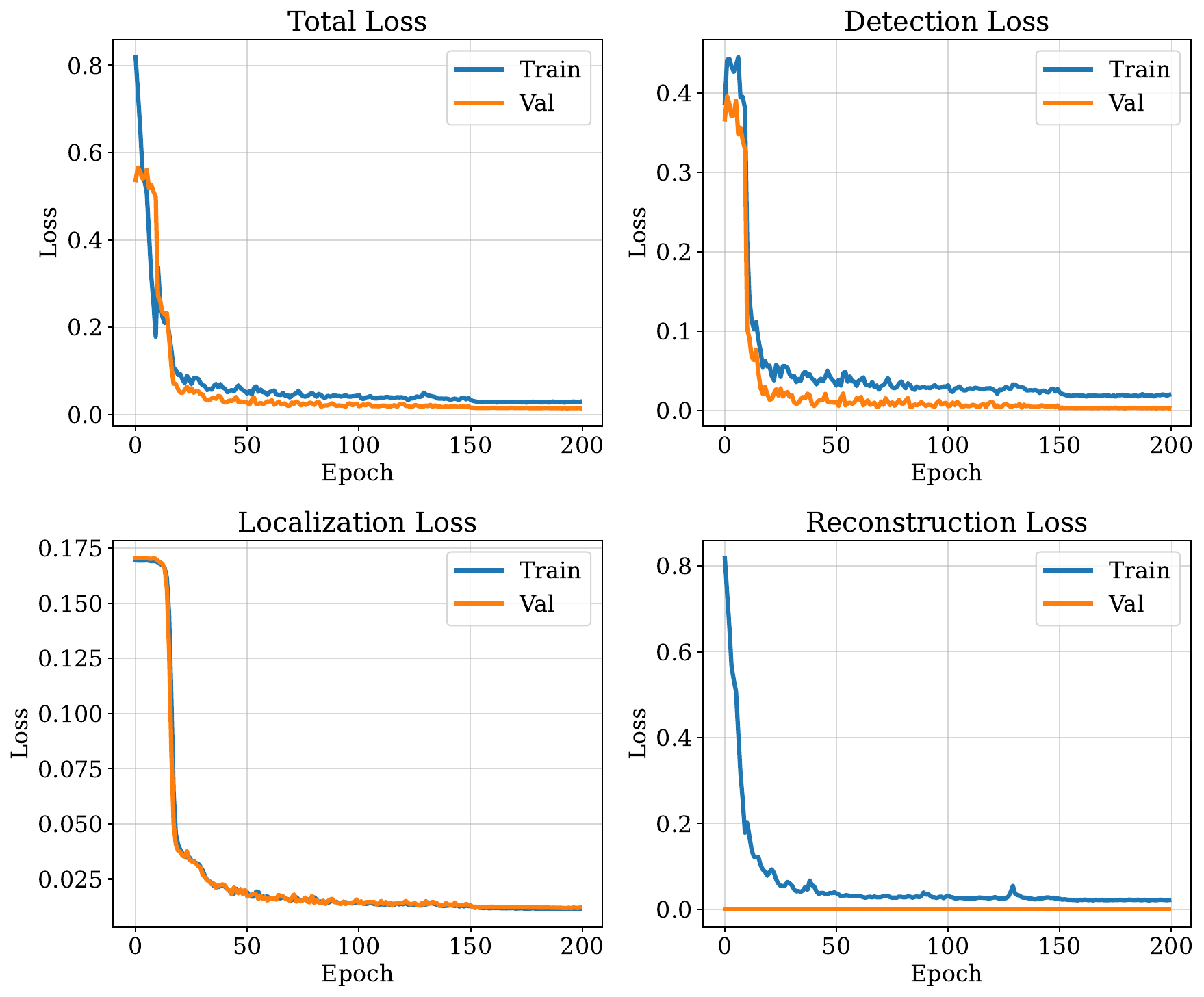}}
    \caption{The training loss curves using data from L03, L04 and L05 specimens that consist of total training loss, HI prediction loss, damage localization loss and reconstruction loss.}
    \label{fig:model_train_loss}
\end{figure}

The proposed data-level Transformer-based fusion method shown in Figure~\ref{fig:data-level fusion} was remotely trained on KU Leuven High Performance Computing (HPC) clusters equipped with NVIDIA A100 80 GB GPUs. A total of 200 epochs were used in the training session. The Adam optimizer was used during the training, with an initial learning rate set to 0.001. A 100-step warm-up strategy was implemented, gradually increasing the learning rate to enhance training stability. The LambdaLR learning rate scheduler was implemented to update the learning rate per epoch using a Lambda function. All training details were logged and visualized on the Weights\&Biases platform. Since the training and validation set segmentation ratio is equal to 4:1, which indicates that $80\%$ of the datasets are set as the training set, and $20\%$ are set as the validation set. Thus, the training and validation loss curves were shown separately in Fig.~\ref{fig:model_train_loss}.

As illustrated in Eq.~\ref{Eq4-9}, the total loss is defined as a weighted linear combination of HI prediction loss, damage localization loss and content reconstruction loss, with coefficients $\lambda_{det}, \lambda_{loc}, \lambda_{rec}$ as 1.0, 0.5, 0.1, respectively. During multi-task training, the model parameters of the prediction head and the localization head are kept frozen for the first 10 epochs, while only the reconstruction head is continuously optimized. This design is motivated by different learning objectives of the task-specific heads. Specifically, the reconstruction head aims to preserve low-level signal characteristics (e.g., frequency, phase, and amplitude information in the PZT signals), whereas the prediction and localization heads are designed to suppress irrelevant raw details and extract task-invariant, damage-related features. 

It should be noted that the reconstruction head is only active during training and is disabled during the validation and testing phases because it mainly functions as a regularizer that shapes the learned representations rather than producing task-specific outputs. Through this two-stage training strategy, the model first learns a meaningful representation of the raw signals at the data level, after which the prediction and localization tasks are jointly optimized. Therefore, both HI prediction and damage localization losses exhibit a significant downward trend from the 10th epoch.

\begin{figure}[H]
    \centering
    \makebox[\textwidth][c]{
    \includegraphics[width=1.1\linewidth]{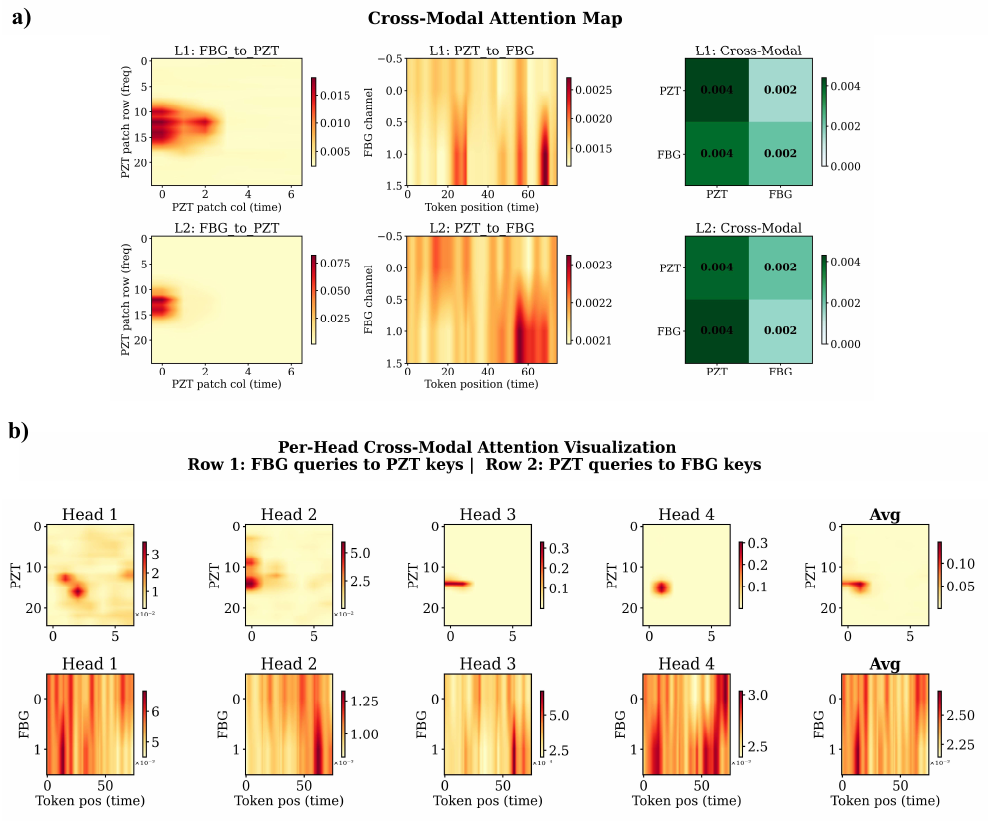}}
    \caption{Attention layer visualization during the Transformer model training process. a) Cross-modal attention map between UGW and FBG modalities; b) Per-head cross-modal attention visualization at the last layer.}
    \label{fig:attn_visual}
\end{figure}

Since the attention module is the core of the Transformer network, Fig.~\ref{fig:attn_visual} visualizes the cross-modal interaction between the PZT and FBG sensor data across two layers (L1 and L2) during training process. The left column displays how FBG queries attend to PZT keys, represented as spectral-temporal patches. The heatmaps indicate that the model prioritizes specific frequency ranges within the ultrasonic signals, particularly between patch rows 10 and 15. In the middle column, ultrasonic queries are seen attending to the FBG channels over time. High-intensity vertical bands at specific token positions (e.g., 20, 45, and 65) indicate that model relies on distinct temporal events in the FBG data to refine its features. The right column provides a summary matrix of the attention weights between these two modalities. At layer 1, the FBG modality exhibits a cross-modal attention weight of 0.004 toward the ultrasonic modality, which is double its self-attention weight of 0.002. This distribution indicates that the network prioritizes inter-sensor correlations over intra-sensor information during fusion. Finally, per-head visualizations reveal how individual heads specialize in capturing distinct cross-modal correlations. The FBG Queries to PZT Keys in Row 1 demonstrate how different heads focus on specific spectral-temporal features, and vice versa in the second row.

\subsubsection{HI prediction results}
\begin{figure}[H]
    \centering
     \makebox[\textwidth][c]{
    \includegraphics[width=0.85\linewidth]{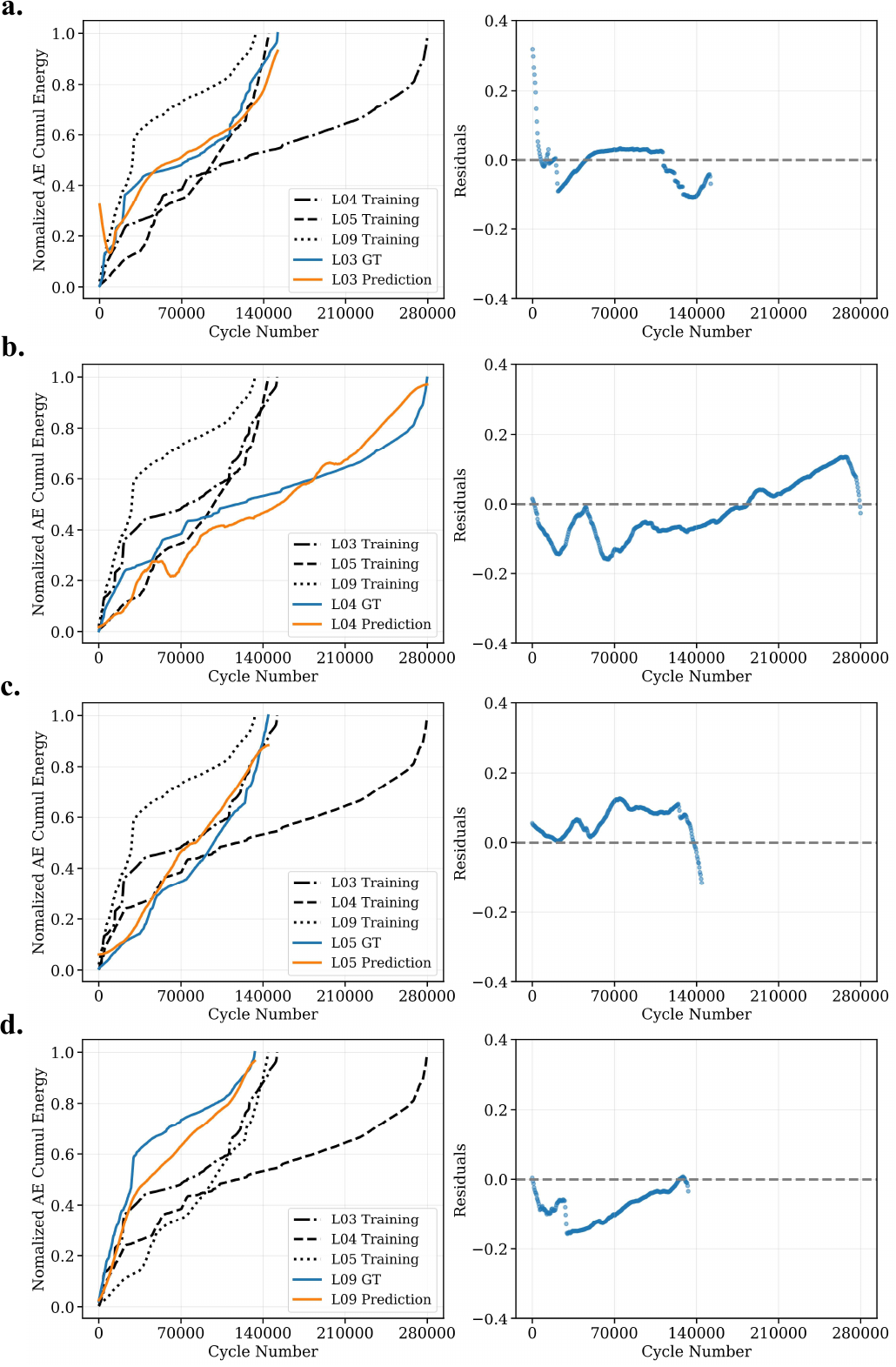}}
    \caption{HI prediction results on L03, L04, L05 and L09 specimens with AE cumulative energy as health indicator.}
    \label{fig:detection results}
\end{figure}

For the HI prediction task on the ReMAP dataset, four composite specimens (L03, L04, L05, and L09) were employed. To comprehensively evaluate model performance under a cross-validation setting, three specimens were used for training and the remaining one for testing in each split. As illustrated in Fig.~\ref{fig:detection results}, the training process employed random sampling in the three training datasets. Since each sample experienced a different number of fatigue loading cycles, only the cumulative AE energy in the annotated data was normalized to the range of 0-1, while the total number of cycles per sample varied.

The left subplot shows that the predicted cumulative AE energy-based HI (orange curve) exhibits an overall increasing trend that is generally consistent with the ground truth. The right subplot presents the residuals between the predicted and actual values, providing a clearer view of model errors. It indicates that prediction accuracy degrades during the mid-fatigue stage where larger deviations occur, whereas the final predictions ultimately converge and closely match the measured values. Notably, the L04 HI curve significantly deviates from the other curves, representing an out-of-distribution (OOD) testing scenario. The fatigue lifetime (cycle number) of L04 is nearly double that of the three training cases (L03, L05, and L09). Despite this substantial discrepancy, the proposed Transformer-based framework is still able to capture the overall HI evolution trend with satisfactory accuracy. This strong generalization capability is mainly attributed to the incorporation of domain-specific prior knowledge into the Transformer training process. 

Since all composite panels share identical geometrical dimensions and material properties, the impact location becomes one of the dominant factors governing the HI evolution behavior. Specifically, the impact position of L04 is located closer to the outer boundary of the plate, whereas the impacts in the remaining specimens are concentrated near the plate center or around the stiffener regions. To implicitly incorporate this physics-informed knowledge, the impact region information is encoded as an additional conditional input to the Transformer model during training. More specifically, the plate surface is partitioned into several predefined spatial regions according to structural characteristics and expected damage sensitivity. Each specimen is then assigned a spatial embedding corresponding to its impact location, which is integrated into the input feature representation. In this way, the Transformer learns not only the temporal evolution patterns of the HI sequence, but also the spatial dependency between impact location and fatigue damage progression.

Furthermore, because damage localization is one of the primary objectives of this study, the impact region is constrained to a localized area with a predefined spatial extent, rather than being treated as a global positional descriptor. This localized regional encoding enables the model to better distinguish the influence of different impact zones on fatigue behavior, thereby improving the robustness and transferability of the learned representation under OOD conditions (for L04 panel).

Based on HI evaluation metrics, Table~\ref{table_4_3} summarizes the quantitative results in terms of MAE, RMSE and $\mathrm{R}^2$. Notably, when L04 is used as the test set and the remaining specimens are used for training, the MAE and RMSE increase to 0.0705 and 0.0808, respectively. This performance degradation can be attributed to the discrepancy between the cumulative AE energy characteristics and fatigue cycling number of specimen L04 and those of L03, L05, and L09. As a result, the features learned during training are more representative of the closed training dataset and generalize poorly to L04. Consistent with this observation, the reduced correlation between the predicted and true HI values causes a drop in the $\mathrm{R^2}$ metric, indicating the challenges of this OOD scenario. However, compared to the PZT-only and FBG-only baselines, the proposed Transformer-based data fusion approach still outperforms them across all test datasets, consistently maintaining an $\mathrm{R^2}$ above 0.8.

\begin{table}[H]
  \centering
  \caption{HI prediction results based on the evaluation metrics.}
  \label{table_4_3}
  \scriptsize
  \renewcommand{\arraystretch}{2}

  \begin{tabular}{m{3cm} m{2.5cm} >{\centering\arraybackslash}m{3cm} m{1.6cm} m{1.6cm} m{1.6cm}}
    \toprule
    \textbf{Methods} & \textbf{Train panels} & \textbf{Test panels} & \textbf{MAE}$\downarrow$ & \textbf{RMSE}$\downarrow$ & $\mathbf{R^2}$$\uparrow$ \\
    \midrule
    \multirow{4}{2cm}{\textbf{PZT-only}} & L04/L05/L09 & L03  & 0.1447 & 0.1705 & 0.3817 \\
    & L03/L05/L09 & L04  & 0.1356 & 0.1570 & 0.3349 \\
      & L03/L04/L09 & L05 & 0.1437 & 0.1602 & 0.5872 \\
        & L03/L04/L05 & L09 & 0.2030 & 0.2339 & -0.0891 \\
    \cmidrule{2-6}
    \multirow{4}{2cm}{\textbf{FBG-only}} & L04/L05/L09 & L03  & 0.2226 & 0.2512 & -0.3415\\
     & L03/L05/L09 & L04  & 0.1531 & 0.1711 & 0.2104  \\
      & L03/L04/L09 & L05  & 0.1631 & 0.1740 & 0.5125 \\
       & L03/L04/L05 & L09  & 0.2283 & 0.2675 & -0.4244 \\
    \cmidrule{2-6}
   \multirow{4}{2cm}{\textbf{Data Fusion method}} & L04/L05/L09 & L03 & \textbf{0.0448} & \textbf{0.0618} & \textbf{0.9188} \\
    & L03/L05/L09 & L04 & \textbf{0.0705} & \textbf{0.0808} & \textbf{0.8236} \\
    & L03/L04/L09 & L05 & \textbf{0.0647} & \textbf{0.0736} & \textbf{0.9130} \\
    & L03/L04/L05 & L09 & \textbf{0.0692} & \textbf{0.0831} & \textbf{0.8887} \\
    \bottomrule
  \end{tabular}
\end{table}

\subsubsection{Damage localization results}
\begin{figure}[H]
    \centering
    \makebox[\textwidth][c]{
    \includegraphics[width=0.95\linewidth]{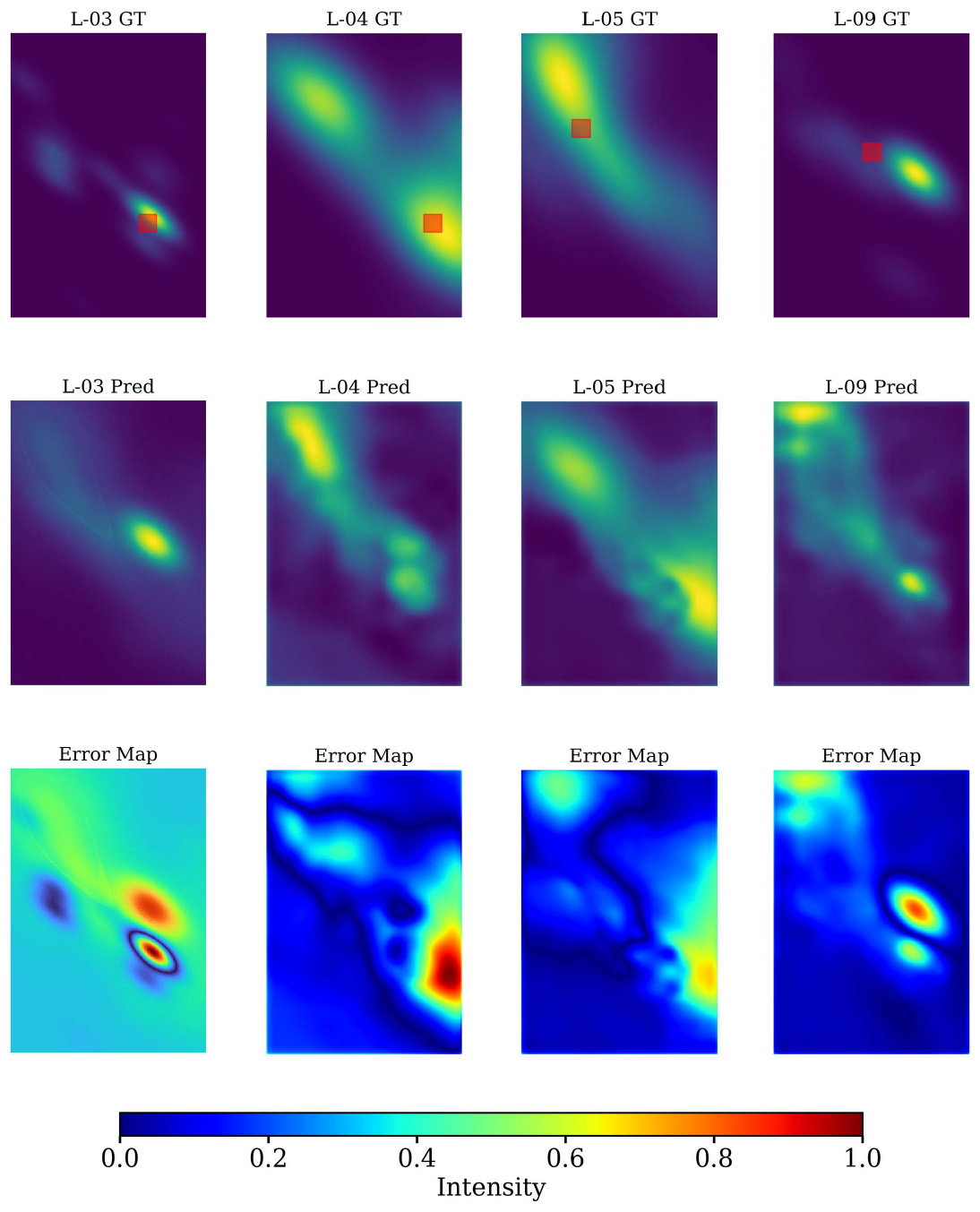}}
    \caption{The damage localization results for composite panels L03, L04, L05, and L09 are presented from left to right columns, respectively. The first row shows the ground truth impact locations obtained from AE localization, with red boxes marking the experimentally validated damage location. The middle row displays the localization results predicted by the Transformer model, and the bottom row represents error maps quantifying the spatial difference between the predicted and actual damage locations.}
    \label{fig:local_pred}
\end{figure}

In addition to the HI prediction task described above, the proposed framework was further evaluated on damage localization performance. Fig.~\ref{fig:local_pred} presents representative localization results for the L09 specimen, including the ground-truth damage maps (top row), the corresponding model predictions (middle row), and the absolute error maps (bottom row) for multiple test cases.

As shown in the first row of Fig.~\ref{fig:local_pred}, the ground truth exhibits locally high-response regions, which indicate the actual damage locations within the composite specimen. The second row displays the damage localization prediction. Overall, the predicted damage distribution accurately captures the primary spatial patterns and locations of the actual damage, demonstrating the model's ability to learn meaningful spatial representations from multimodal sensor data. Notably, the regions with enhanced predicted responses typically align closely with the annotated damage locations, confirming the effectiveness of the proposed localization strategy.

The third row of Fig.~\ref{fig:local_pred} visualizes the absolute error between the predicted and target maps, providing a more intuitive assessment of localization accuracy. The error map reveals that most deviations are concentrated near the boundaries of the damaged region, while reconstruction errors in the central damaged area are relatively low. This phenomenon indicates that the model excels at identifying the approximate location of damage rather than precisely delineating its spatial extent. Nevertheless, the overall error magnitude remains within a limited range, indicating stable and consistent localization performance across different targets.

Table~\ref{table_4_4} illustrates the quantitative damage localization performance in terms of MAE, RMSE, and SSIM under different training and testing settings between samples. Overall, the proposed model achieves relatively low MAE and RMSE values across all test cases, indicating that the predicted damage maps are numerically close to the corresponding ground-truth distributions.

When L04 is used as the OOD test panel, the model produces the highest MAE and RMSE values (0.0465 and 1571, respectively), suggesting comparatively larger localization errors. This outcome can be attributed to the distinct damage evolution patterns and spatial characteristics of L04 relative to the training specimens (L03, L05, and L09). In contrast, testing on L03 and L05 results in notably improved performance, with lower MAE and RMSE values and the highest SSIM scores (0.8892 and 0.6426, respectively). The higher SSIM values indicate that the output localization maps for these specimens not only match the ground truth in magnitude but also preserve the underlying structural and spatial similarity. For the L09 panel, although MAE and RMSE remain comparable to those of L05, the SSIM value decreases to 0.6115. This suggests that while the overall error magnitude is limited, the spatial distribution of the predicted damage differs more significantly from the ground truth. This observation is consistent with the qualitative results shown in Fig.~\ref{fig:local_pred}, where the predicted regions for the L09 panel capture the general damage location but exhibit reduced structural consistency compared to other panels.

\begin{table}[H]
  \centering
  \caption{Damage localization results based on the evaluation metrics.}
  \label{table_4_4}
  \scriptsize
  \renewcommand{\arraystretch}{2}

  \begin{tabular}{m{3cm} m{2.5cm} >{\centering\arraybackslash}m{3cm} m{1.6cm} m{1.6cm} m{1.6cm}}
    \toprule
    \textbf{Methods} & \textbf{Train panels} & \textbf{Test panels} & \textbf{MAE}$\downarrow$ & \textbf{RMSE}$\downarrow$ & \textbf{SSIM}$\uparrow$ \\
    \midrule
    \multirow{4}{2cm}{\textbf{PZT-only}} & L04/L05/L09 & L03  & 0.0601 & 0.2006 & 0.3500 \\
    & L03/L05/L09 & L04  & 0.0502 & 0.1547 & 0.5486 \\
      & L03/L04/L09 & L05 & 0.0559 & 0.1706 & 0.4615 \\
        & L03/L04/L05 & L09 & 0.0895 & 0.2248 & 0.1517 \\
    \cmidrule{2-6}
    \multirow{4}{2cm}{\textbf{FBG-only}} & L04/L05/L09 & L03  & 0.1089 & 0.2691 & 0.2555 \\
     & L03/L05/L09 & L04 & 0.0559 & 0.1570 & 0.3331  \\
      & L03/L04/L09 & L05 & 0.0806 & 0.1790 & 0.4573 \\
       & L03/L04/L05 & L09 & 0.1025 & 0.2330 & 0.1891 \\
    \cmidrule{2-6}
    \multirow{4}{2cm}{\textbf{Data Fusion method}} & L04/L05/L09 & L03 & \textbf{0.0400} & \textbf{0.1520} & \textbf{0.8892} \\
    & L03/L05/L09 & L04 & \textbf{0.0465} & \textbf{0.1571} & \textbf{0.6072} \\
    & L03/L04/L09 & L05 & \textbf{0.0384} & \textbf{0.1444} & \textbf{0.6426} \\
    & L03/L04/L05 & L09 & \textbf{0.0455} & \textbf{0.1545} & \textbf{0.6115} \\
    \bottomrule
  \end{tabular}
\end{table}

\subsubsection{Transformer-based fusion vs State-of-the-art DNN models}
To further validate the proposed Transformer-based data fusion scheme, we compared its performance against state-of-the-art (SOTA) deep learning architectures, specifically 1D-CNN, Bi-LSTM, and CNN-LSTM. To ensure a consistent baseline, all models share the same tokenization pipeline, including the tokenizer and data loading process. Transformer-based models contain significantly more parameters than SOTA DNN models, which enables them with a powerful capability to process multi-sensor data streams and capture nonlinear degradation trends. However, as mentioned earlier, model inference efficiency should be considered during the data fusion process. 

As shown in Table~\ref{table_4_5}, it gives the inference times for the Transformer, 1D-CNN, Bi-LSTM and CNN-LSTM models. The CNN models achieve the fastest performance, with an average inference time of 0.18s across all test groups. This speed stems from the highly parallelizable nature of local convolution operations, which are heavily optimized for the tensor cores of modern hardware (such as the NVIDIA A100 GPU utilized in this study). In contrast, the Bi-LSTM model exhibited a noticeably slower baseline inference time (around 0.25s), hindered by the strictly sequential dependency of its recurrent operations. The hybrid CNN-LSTM model effectively split the difference, with its inference speed falling directly between the 1D-CNN and Bi-LSTM. 

Notably, a distinct behavior was observed on the L04 dataset, where the test data nearly doubles compared to the L03, L05, and L09 panels. Under these circumstances, the inference time of the Transformer model nearly doubled, scaling aggressively with test data length due to the inherent $\mathcal{O}(N^2)$ quadratic complexity of its self-attention mechanism. By contrast, the 1D-CNN, Bi-LSTM, and CNN-LSTM models demonstrated remarkable temporal resilience, showing almost no slowdown on the longer L04 test set. Because these architectures operate with $\mathcal{O}(N)$ linear time complexity, their computational profiles at these scales are dominated by fixed framework overhead and GPU kernel launch latencies.

\begin{table}[H]
  \centering
  \caption{Model inference time (unit: seconds) comparison between the proposed Transformer-based data-level fusion model and SOTA DNN models using the ReMAP dataset.}
  \label{table_4_5}
  \scriptsize
  \renewcommand{\arraystretch}{2.5}

  \begin{tabular}{>{\centering\arraybackslash}m{3.5cm} >{\centering\arraybackslash}m{1.5cm} >{\centering\arraybackslash}m{1.5cm} >{\centering\arraybackslash}m{1.5cm} >{\centering\arraybackslash}m{1.5cm} }
    \toprule
    \diagbox[width=3.5cm]{\textbf{Models}}{\textbf{Panels}} & \textbf{L03} & \textbf{L04} & \textbf{L05} & \textbf{L09} \\
    \midrule
    Transformer & \textbf{0.2940} & \textbf{0.5356} & \textbf{0.2822} & \textbf{0.2759} \\
    1D-CNN & 0.1878 & 0.1997 & 0.1821 & 0.1793 \\
    Bi-LSTM & 0.2425 & 0.3139 & 0.2563 & 0.2644 \\
    CNN-LSTM & 0.2406 & 0.2691 & 0.2394 & 0.2470 \\
     \bottomrule
\end{tabular}
\end{table}

\begin{figure}[H]
    \centering
    \includegraphics[width=0.7\linewidth]{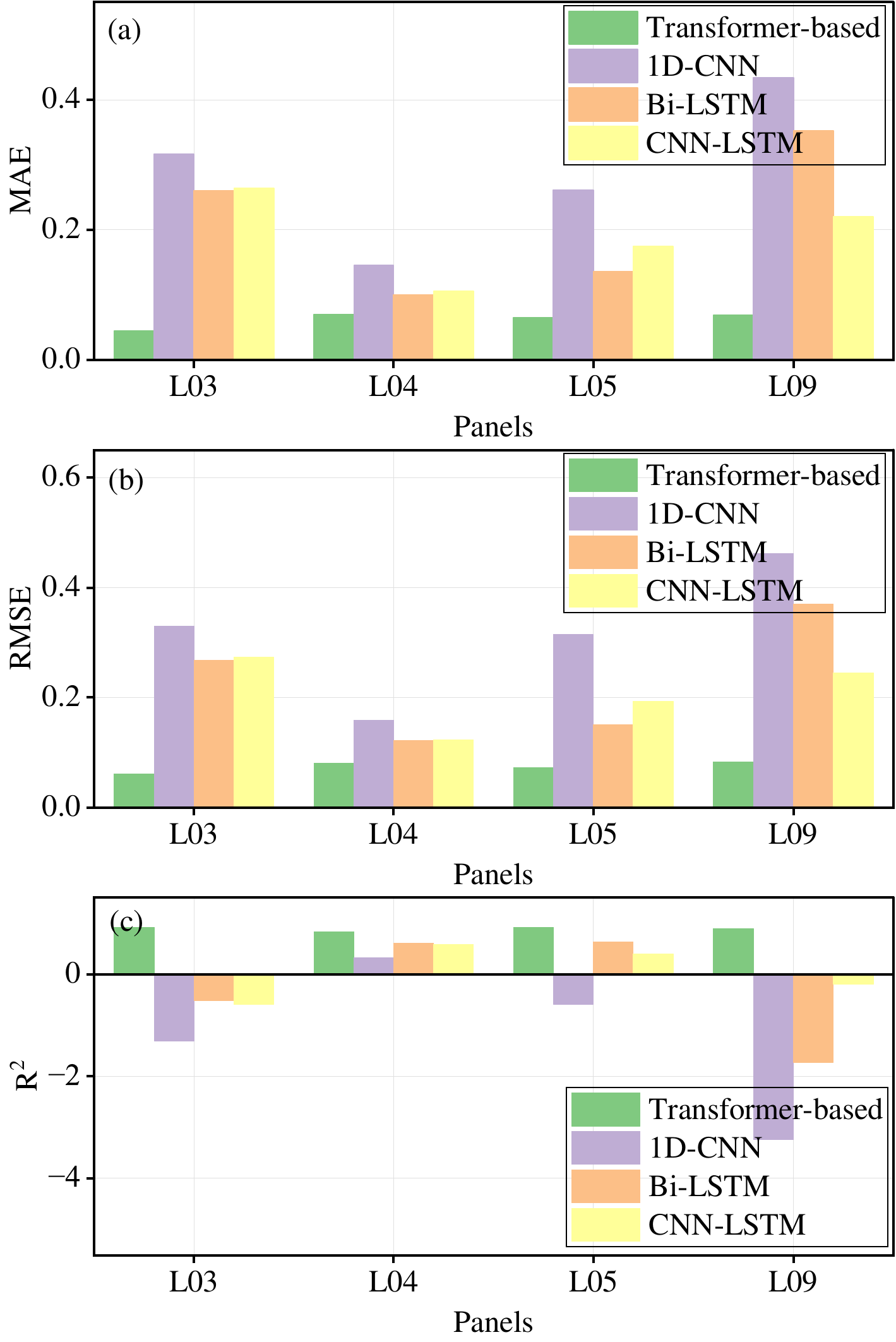}
    \caption{A comparison of HI prediction results between the proposed Transformer-based data fusion model and SOTA DNN models. (a) MAE value; (b) RMSE value; (c) $\mathrm{R}^2$ values.}
    \label{fig:sota_compare_datalevel}
\end{figure}

However, as shown in Fig.~\ref{fig:sota_compare_datalevel}, the proposed Transformer-based data fusion approach achieves the best performance across MAE, RMSE, and $\mathrm{R}^2$ metrics compared to other SOTA DNN models. Among the benchmarked architectures, the CNN-LSTM model outperforms both the 1D-CNN and Bi-LSTM, yielding lower MAE and RMSE values. This can be attributed to the hybrid architecture of the CNN-LSTM model to leverage the strengths of both spatial feature extraction and temporal modeling. The 1D-CNN model, on the other hand, exhibits the poorest performance across all evaluation metrics, suggesting that a simple 1D-CNN is insufficient for sophisticated multi-sensor data fusion tasks.

In summary, while the proposed Transformer-based data fusion scheme achieves the highest HI prediction and damage localization accuracy, this comes at the cost of slower inference speeds. For industrial applications, the choice of deployment ultimately depends on whether the operator prioritizes HI prediction and localization accuracy or the model’s real-time efficiency.

\section{Conclusion}
\label{Sec:5}
This paper proposes a Transformer-based data fusion framework and validates it using a ReMAP composite panel as an example. The proposed Transformer network takes multi-channel piezoelectric transducer (PZT) and fiber Bragg grating (FBG) sensor data as input and yields multi-task outputs: prediction of health indicator (HI) progression and damage localization. Since the sampling frequency and acquisition interval of ultrasonic guided wave signals and strain data differ, the Transformer network fuses the data from these two types of sensors by slicing the original measurments into tokens and synchronously fusing the fused tokens. In this way, the proposed Transformer network can handle the differences in sampling frequency and acquisition intervals between different sensor modalities.

In experimental validation, the proposed Transformer-based data fusion framework achieves superior performance in HI prediction compared to using PZT or FBG data alone. Across all test samples, the mean absolute error (MAE) and root mean squared error (RMSE) remained below 0.0831, while the coefficient of determination ($\mathrm{R}^2$) exceeded 0.8, even under the out-of-distribution (OOD) test scenario for the L04 panel. For damage localization, the model achieved the lowest error metrics, with MAE and RMSE values below 0.05 and 0.16, respectively, and the highest structural similarity index measure, exceeding 0.6.  These results validate the effectiveness of the proposed Transformer-based data fusion model. Compared with state-of-the-art (SOTA) DNN models, the proposed Transformer-based model doubles the inference time on OOD test set, yet still achieves the highest scores across all evaluation metrics. In conclusion, by appropriately reducing model inference efficiency while accounting for industrial applications, the proposed method can outperform benchmark single-source Transformer model and SOTA DNN models.

\section{Authorship contribution statement}
\textbf{Xin Yang:} Conceptualization, Methodology, Software, Visualization, Validation, Writing - Original Draft, Writing - Review \& Editing.  
\textbf{Morteza Moradi:} Conceptualization, Resources, Writing - Review \& Editing.  \textbf{Tongtong Yan:} Methodology, Writing - Review \& Editing. \textbf{Jinbo Du:} Methodology, Writing - Review \& Editing. \textbf{Yunlai Liao:}  Methodology, Writing - Review \& Editing. \textbf{Dimitrios Zarouchas:} Resources, Writing - Review \& Editing. \textbf{Dimitrios Chronopoulos:} Supervision, Project administration, Writing - Review \& Editing.

\clearpage
\bibliography{Refs}

\end{document}